\documentclass[twocolumn,superscriptaddress,letter]{revtex4}%
\usepackage{amssymb}
\usepackage{color}
\usepackage{graphicx}
\usepackage{dcolumn}
\usepackage{bm}
\usepackage[header,title,page,titletoc]{appendix}
\usepackage{amsmath}
\usepackage{amsfonts}%
\providecommand{\U}[1]{\protect \rule{.1in}{.1in}}
\begin{document}
\title{Non-Hermitian Dynamic Strings and Anomalous Topological Degeneracy on
non-Hermitian Toric-code Model with Parity-time Symmetry}
\author{Cui-Xian Guo}
\thanks{These authors contributed equally to the work}
\affiliation{Center for Advanced Quantum Studies, Department of Physics, Beijing Normal
University, Beijing 100875, China}
\author{Xiao-Ran Wang}
\thanks{These authors contributed equally to the work}
\affiliation{Center for Advanced Quantum Studies, Department of Physics, Beijing Normal
University, Beijing 100875, China}
\author{Can Wang}
\affiliation{Center for Advanced Quantum Studies, Department of Physics, Beijing Normal
University, Beijing 100875, China}
\author{Su-Peng Kou}
\thanks{Corresponding author}
\email{spkou@bnu.edu.cn}
\affiliation{Center for Advanced Quantum Studies, Department of Physics, Beijing Normal
University, Beijing 100875, China}

\begin{abstract}
In this paper, with the help of Hermitian/non-Hermitian dynamic strings, the
theory of non-Hermitian topological order is developed based on a
non-Hermitian Wen-plaquette model. The effective models for bosonic
topological excitations (e-particle and m-particle) are Hermitian
tight-binding lattice model; the effective model for fermionic topological
excitation (f-particle) becomes a non-Hermitian tight-binding lattice model.
In addition, the effective pseudo-spin model for topologically degenerate
ground states is derived by calculating the expectation values of
Hermitian/non-Hermitian topological closed dynamic strings. For the
topologically degenerate ground states of non-Hermitian Wen-plaquette model on
an even-by-odd, odd-by-even and odd-by-odd lattice, anomalous topological
degeneracy occurs, i.e., the number of the topologically protected ground
states may be reduced from $2$ to $1$. Now, the effective pseudo-spin model
turns into the typical $\mathcal{PT}$-symmetric non-Hermitian Hamiltonian with
spontaneous $\mathcal{PT}$-symmetry breaking. At \textquotedblleft exceptional
points\textquotedblright, the topologically degenerate ground states merge
with each other and the topological degeneracy turns into non-Hermitian
degeneracy. In the end, the application of the non-Hermitian $Z_{2}$
topological order and its possible physics realization are discussed.

\end{abstract}

\maketitle

\section{Introduction}

The non-Hermitian Hamiltonians obeying Parity-time ($\mathcal{PT}$)-symmetry
as a complex extension of the (Hermitian) quantum mechanics can exist purely
real energy spectrum, which was proposed by Bender and Boettcher\cite{Bender
98}. It can exists spontaneous $\mathcal{PT}$-symmetry breaking accompanied by
spectral transition from real to complex. The energy degeneracy points in
non-Hermitian system are called \textquotedblleft \emph{exceptional
points}\textquotedblright \ (EPs), at which energies coalesce and eigenvectors
merge with each other. In physics, $\mathcal{PT}$-symmetric non-Hermitian
systems are always obtained by appropriately engineered gain and loss. There
are a variety of suitable platforms to realize $\mathcal{PT}$-symmetric
non-Hermitian systems, such as
optics\cite{Guo2009,Ruter2010,Chong2011,Regensburger2012,Feng2013,Wimmer2015,Weimann2017,Hodaei2017,Ozdemir2019}%
, electronics\cite{Schindler2011,Assawaworrarit2017,Choi2018},
microwaves\cite{Bittner2012,Peng2014,Poli2015,Liu2016},
acoustics\cite{ZhuX2014,Popa2014,Fleury2015,Ding2016}, single-spin
system\cite{Wu2019}, and dynamic systems\cite{Xiao2017,luo,Xiao2019,Wang2019}.
Some applications associated with $\mathcal{PT}$-symmetric system have been
explored including unidirectional transport\cite{Feng2013,Peng2014} and
single-mode lasers\cite{Feng2014,Hodaei2014}.

On the other hand, the non-Hermitian topological states and topological
invariants of tight-binding models have been widely
studied\cite{Rudner2009,Esaki2011,Hu2011,Liang2013,Zhu2014,Lee2016,San2016,Leykam2017,Shen2018,Lieu2018,
Xiong2018,Kawabata2018,Gong2018,Yao2018,YaoWang2018,Yin2018,Kunst2018,KawabataUeda2018,Alvarez2018,
Jiang2018,Ghatak2019,Avila2019,Jin2019,Lee2019,Liu2019,38-1,38,chen-class2019,Edvardsson2019,
Herviou2019,Yokomizo2019,zhouBin2019,Kunst2019,Deng2019,SongWang2019,xi2019,Longhi2019,chen-edge2019}%
. The classification of non-Hermitian topological phases in terms of
symmetries has been developed including $\mathcal{PT}$%
-symmetry\cite{Gong2018,38-1,38}. However, there exists another type of
topological states beyond Landau's symmetry breaking paradigm -- topological
orders\cite{wen3}. For the topological orders, all the excitations are gapped
and have fractional statistics and the ground states have topological
degeneracy, i.e., the quantum degeneracy of the ground states depends on the
genius of the manifold of the
background\cite{wen3,k1,k2,wen1,wen2,kou1,kou2,kou3,md}. The degenerate ground
states of a topological order (on a torus) make up a protected code subspace
(the topological qubits). {It is possible to incorporate intrinsic fault
tolerance into a quantum computer - topological quantum computation (TQC)
which} avoids decoherence and is free from errors\cite{k1,k2}.

A question is "\emph{How about topological orders meet }$\mathcal{PT}%
$\emph{-symmetric non-Hermitian?}" In this paper, by taking the non-Hermitian
Wen-plaquette (toric-code) model as an example, we will study a $Z_{2}$
topological order on a $\mathcal{PT}$ symmetric non-Hermitian system and
develop a theory for non-Hermitian ($Z_{2}$) topological order. The quantum
states of non-Hermitian Wen-plaquette (toric-code) are characterized by
Hermitian/non-Hermitian dynamic strings. The topological degeneracy for the
system on even-by-odd, odd-by-even and odd-by-odd lattices becomes anomalous:
the topologically degenerate ground states may merge with each other and the
topological degeneracy turns into non-Hermitian degeneracy. The number of the
topologically protected ground states turns to $1$ rather than $2$. We call it
anomalous topological degeneracy.

In this paper, we will develop a theory of the non-Hermitian $Z_{2}$
topological order by taking the Wen-plaquette model as an example. The
remainder of the paper is organized as follows. In Sec. II, we review the
theory of Hermitian $Z_{2}$ topological order (the Wen-plaquette model). In
Sec. III, a non-Hermitian Wen-plaquette model is proposed. In Sec. IV, based
on biorthogonal set, the theory for non-Hermitian quantum string states in the
non-Hermitian $Z_{2}$ topological order is developed. In Sec. V, the
propertities of quasi-particles in the non-Hermitian $Z_{2}$ topological order
is given. The effective model for fermionic quasi-particles is non-Hermitian
and the energy spectra become complex. In Sec. VI, the physics of anomalous
topological degeneracy are explored. The effective model for the degenerate
ground states on even-by-odd, odd-by-even and odd-by-odd lattices is derived
that always has $\mathcal{PT}$-symmetry. In this section, spontaneous
$\mathcal{PT}$-symmetry breaking for the topologically protected degenerate
ground states is discovered. The numerical results confirm the analytic
theoretical predictions. Finally, the conclusions are given in Sec. VII.

\section{$Z_{2}$ topological order for Hermitian Wen-plaquette model}

Firstly, we review the topological properties of (Hermitian) $Z_{2}$
topological order. There are several exactly solvable spin models with $Z_{2}$
topological orders, such as the Kitaev toric-code model \cite{k1}, the
Wen-plaquette model \cite{wen1,wen2} and the Kitaev model on a honeycomb
lattice \cite{k2}. In this paper, we take Wen-plaquette model as an example.

\subsection{The (Hermitian) Wen-plaquette model}

The Hamiltonian of the (Hermitian) Wen-plaquette
model\cite{wen3,k1,k2,wen1,wen2} is
\begin{equation}
\hat{H}=-g\sum_{i}\hat{F}_{i},
\end{equation}
with
\begin{equation}
F_{i}=\sigma_{i}^{x}\sigma_{i+\hat{e}_{x}}^{y}\sigma_{i+\hat{e}_{x}+\hat
{e}_{y}}^{x}\sigma_{i+\hat{e}_{y}}^{y},
\end{equation}
and $g>0.$ $\sigma_{i}^{x},$ $\sigma_{i}^{y}$ are Pauli matrices on site $i$.
In this paper, $g$\ is set to be unit, $g\equiv1$. In Fig.1, we show the
schematic diagram of the (Hermitian) Wen-plaquette model. The ground state is
a topological order described by $Z_{2}$ gauge symmetry that is denoted by
${F_{i}\equiv+1}$ at each plaquette with the ground state energy
\begin{equation}
E_{0}=-g\mathcal{N},
\end{equation}
where $\mathcal{N}$ is the total lattice number\cite{wen1,wen2}.

\begin{figure}[ptb]
\includegraphics[clip,width=0.3\textwidth]{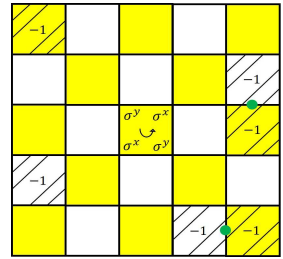}\caption{(Color online)
The schematic diagram of the (Hermitian) Wen-plaquette model. The yellow/white
plaquette stands for odd/even sub-plaquette, respectively. The shadow region
in odd/even sub-plaquette represent an m-particle and e-particle,
respectively. Green spheres on the links represent f-particles.}%
\end{figure}

\subsection{(Hermitian) quantum string states}

The quantum states of the (Hermitian) Wen-plaquette model are characterized by
different configurations of strings\cite{wen3,k1,k2,wen1,wen2},
\begin{equation}
\left \vert \Phi \right \rangle =\sum_{\mathcal{C}}a_{\mathcal{C}}\hat
{W}(\mathcal{C})|0\rangle,
\end{equation}
where $\hat{W}(\mathcal{C})=\prod_{\mathcal{C}}\sigma_{i}^{\alpha_{i}}$ denote
the possible open/closed string operators, and $a_{\mathcal{C}}$\ is weight of
the string operator. $\sigma_{i}^{\alpha_{i}}$ is $\alpha_{i}$-type Pauli
matrix on site $i$ and $\prod_{i\in \mathcal{C}}$ is over all the sites on the
string along a loop $\mathcal{C}$, and, where $|0\rangle$\ denotes the spin
polarized states with all spin down ($\left \vert \downarrow \downarrow
,\cdots,\downarrow \right \rangle $). Here, the string operators are all
Hermitian, i.e.,
\begin{equation}
\hat{W}(\mathcal{C})=\hat{W}^{\dagger}(\mathcal{C}),
\end{equation}
and for quantum string states we have
\begin{equation}
\hat{W}(\mathcal{C})\left \vert \Phi \right \rangle =\hat{W}^{\dagger
}(\mathcal{C})\left \vert \Phi \right \rangle =d(\mathcal{C})\left \vert
\Phi \right \rangle ,
\end{equation}
where $d(\mathcal{C})$ is eigenvalue and $\left \vert \Phi \right \rangle $ is
the eigenstate. For example, the ground state is an equal superposition of
loop (closed string) configurations: $\sum_{\mathcal{C}\in
\mathrm{closed-string}}\hat{W}(\mathcal{C})|0\rangle$. The closed loops are
interpreted as electric field lines of the $Z_{2}$ gauge theory. On a torus,
the ground states have topological degeneracy, of which the wave-functions are
characterized by topological closed strings.

\subsection{Energy spectra of quasi-particles}

In the Wen-plaquette model, there are three types of quasi-particles,
e-particle (or $Z_{2}$ charge), m-particle (or $Z_{2}$ vortex), and f-particle
(or fermion). m-particle is defined as ${F_{i}=-1}$ at odd sub-plaquette and
e-particle is defined as ${F_{i}=-1}$ at even sub-plaquette. The mass gap of
e-particle and m-particle is $2g$. In particular, the bound states of an
e-particle and an m-particle on two neighbor plaquettes are f-particle obeying
fermionic statistic. See the illustration in Fig.1. All quasi-particles in
such exactly solvable model can't move. The energy spectra are $E_{v}%
=E_{c}=2g$ for e-particle and m-particle, $E_{f}=4g$ for fermion, respectively.

These three types of quasi-particles (e-particle, m-particle, and f-particle)
can be characterized by three types of (open) string operators $W_{c}%
(\mathcal{C})$, $W_{v}(\mathcal{C})$, $W_{f}(\mathcal{C})$ that are all
Hermitian,
\begin{align}
W_{c}(\mathcal{C})  &  =W_{c}^{\dagger}(\mathcal{C}),\text{ }W_{v}%
(\mathcal{C})=W_{v}^{\dagger}(\mathcal{C}),\nonumber \\
W_{f}(\mathcal{C})  &  =W_{f}^{\dagger}(\mathcal{C}).
\end{align}
For each open string, the corresponding quantum states are two excited
quasi-particles. The string operator for a pair of e-particles $W_{c}%
(\mathcal{C})=\prod_{\mathcal{C}}\sigma_{i}^{\alpha_{i}}$ (or a pair of
m-particles $W_{v}(\mathcal{C})=\prod_{\tilde{\mathcal{C}}}\sigma_{i}%
^{\alpha_{i}}$) is the product of spin operators along a loop $\mathcal{C}$
(or $\tilde{\mathcal{C}}$) connecting adjacent even-plaquettes (or
odd-plaquettes), $\alpha_{i}=y$ if $i$ is even and $\alpha_{i}=x$ if $i$ is
odd. The string for a pair of f-particles described by the $W_{f}%
(\mathcal{C})=W_{c}(\mathcal{C})W_{v}(\mathcal{C})$ can be regarded as a
combination of the string for e-particles and m-particles.

It was known that for the Wen-plaquette model, the quasi-particles cannot move
in the solvable limit. The situation changes after adding external fields.
Under the perturbation $H_{I}=h_{x}\sum \limits_{i}\sigma_{i}^{x}+h_{z}%
\sum \limits_{i}\sigma_{i}^{z},$ the quasi-particles (e-particle, m-particle
and f-particle) begin to hop. The term $h_{x}\sum \limits_{i}\sigma_{i}^{x}$
drives the e-particle, m-particle and f-particle hopping along diagonal
direction $\hat{e}_{x}-\hat{e}_{y}$, and the term $h_{z}\sum \limits_{i}%
\sigma_{i}^{z}$ drive f-particle hopping along $\hat{e}_{x}$ or $\hat{e}_{y}$
direction without affecting e-particle and m-particle. There exist two types
of f-particles: the f-particles on the vertical links and the f-particles on
the parallel links. The term $h_{z}\sum \limits_{i}\sigma_{i}^{z}$ drives the
f-particles on the vertical links move along vertical directions and the
f-particles on the parallel links move along parallel directions. That means
both types of f-particle cannot turn round any more under the term $h_{z}%
\sum \limits_{i}\sigma_{i}^{z}$. But these two types of f-particles can be
converted into each other under the term $h_{x}\sum \limits_{i}\sigma_{i}^{x}$,
as has been shown in Fig.2.

\begin{figure}[ptb]
\includegraphics[clip,width=0.42\textwidth]{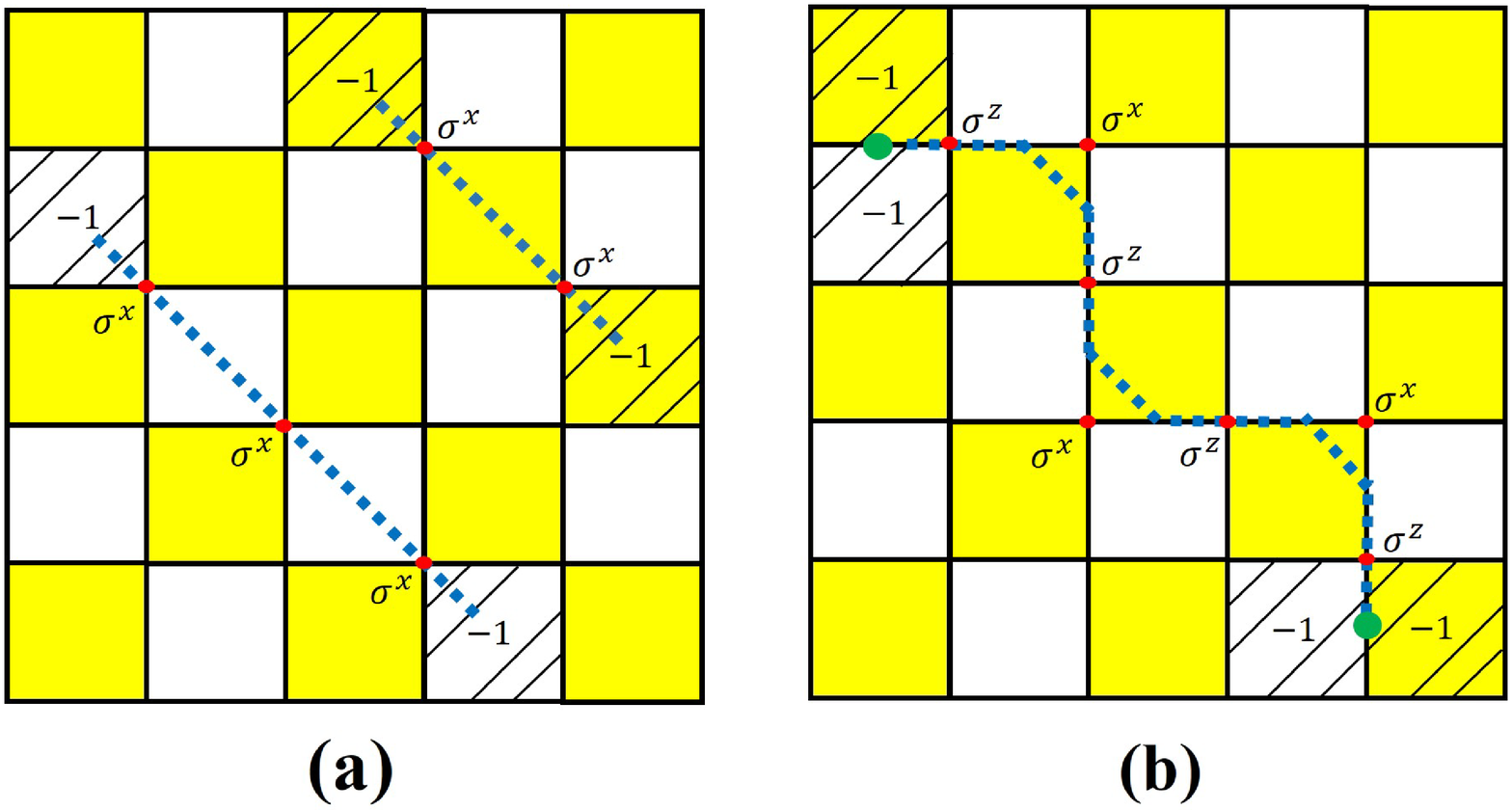}\caption{(Color online)
(a) The schematic diagram of strings for e/m-particle under the term $\hat
{H}_{I}$; (b) The schematic diagram of strings for f-particle under the term
$\hat{H}_{I}$.}%
\end{figure}

Let us calculate the energy spectra for three types of quasi-particles.

For an e-particle (or m-particle) living at $i$ plaquette ${F_{i}=-1,}$ when
$\sigma^{x}$ acts on $i+\hat{e}_{x}$ site, it hops to $i+\hat{e}_{x}-\hat
{e}_{y}$ plaquette denoted by ${F_{i+\hat{e}_{x}-\hat{e}_{y}}=-1}$. Moreover,
a pair of e-particles (or m-particles) located at $i$ and $i+\hat{e}_{x}%
-\hat{e}_{y}$ plaquettes can be created by the operator of $\sigma_{i+\hat
{e}_{x}}^{x}$,
\begin{align}
{F_{i}}  &  {=}{+1\rightarrow F_{i}=-1,}\nonumber \\
{F_{i+\hat{e}_{x}-\hat{e}_{y}}}  &  {=}{+1\rightarrow F_{i+\hat{e}_{x}-\hat
{e}_{y}}=-1},
\end{align}
or annihilated also
\begin{align}
{F_{i}}  &  {=}{-1\rightarrow F_{i}=+1,}\nonumber \\
F{_{i+\hat{e}_{x}-\hat{e}_{y}}}  &  {=}{-1\rightarrow F_{i+\hat{e}_{x}-\hat
{e}_{y}}=+1.}%
\end{align}
With the help of perturbative method, each hopping term $\hat{t}_{I^{a}J^{a}%
}^{a}$ for e/m-particle on $I^{a}$-lattice ($a=v$ on lattice of odd
sub-plaquettes and $a=c$ on lattice of even sub-plaquettes) corresponds to a
one-step Hermitian string operator $\hat{t}_{i}^{a}=h_{x}\sigma_{i}^{x}$,
\begin{align}
\hat{t}_{i}^{a}  &  =h_{x}\sigma_{i}^{x}\rightarrow \nonumber \\
\hat{t}_{I^{a}J^{a}} ^{a}  &  =h_{x}(\phi_{a,I^{a}}^{\dag}\phi_{a,I^{a}%
+e_{x^{I^{a}}}^{I^{a}}} +\phi_{a,I^{a}}^{\dag}\phi_{a,I^{a}+e_{x^{I^{a}}%
}^{I^{a}}}^{\dag}+h.c.),
\end{align}
where $\phi_{a,I^{a}}^{\dag}$ is the generation operator of the e/m-particle.
Here, $e_{x^{I^{a}}}^{I^{a}}$ is the unit vector along $x^{I^{a}}$ direction
on $I^{a}$-lattice, of which the lattice constant is $\sqrt{2}a_{0}$ ($a_{0}$
is lattice constant of the original square lattice). See the illustration in Fig.3(a).

The effective Hamiltonian of boson $\hat{H}_{\mathrm{eff}}^{a}$ ($a=v$ denote
m-particle, $a=c$ denote e-particle) on $L_{x}\times L_{y}$ square lattice can
be expressed as
\begin{align}
\hat{H}_{\mathrm{eff}}^{a}=  &  \sum_{\left \langle I^{a},J^{a}\right \rangle
}h_{x}(\phi_{a,I}^{\dag}\phi_{a,I^{a}+e_{x^{I^{a}}}^{I^{a}}}+\phi_{a,I}^{\dag
}\phi_{a,I^{a}+e_{x^{I^{a}}}^{I^{a}}}^{\dag}+h.c.)\nonumber \\
&  +2g\sum_{I^{a}}\phi_{a,I^{a}}^{\dag}\phi_{a,I^{a}}.
\end{align}
After Fourier transformation, $\hat{H}_{\mathrm{eff}}^{a}$ is written as
\begin{equation}
\hat{H}_{\mathrm{eff}}^{a}=\sum_{k_{x}^{a}}\left(
\begin{array}
[c]{cc}%
\phi_{k_{x}^{a}}^{\dagger} & \phi_{-k_{x}^{a}}%
\end{array}
\right)  \mathcal{\hat{H}}_{\mathrm{eff}}^{a}\left(
\begin{array}
[c]{c}%
\phi_{k_{x}^{a}}\\
\phi_{-k_{x}^{a}}^{\dagger}%
\end{array}
\right)  -(2g+2h_{x}\cos k_{x}^{a}),
\end{equation}
where $\mathcal{\hat{H}}_{\mathrm{eff}}^{a}$ in momentum space is obtained as
\begin{equation}
\mathcal{\hat{H}}_{\mathrm{eff}}^{a}=\left(
\begin{array}
[c]{cc}%
2g+2h_{x}\cos k_{x}^{a} & 2h_{x}\cos k_{x}^{a}\\
2h_{x}\cos k_{x}^{a} & 2g+2h_{x}\cos k_{x}^{a}%
\end{array}
\right)  .
\end{equation}

Then we use Bogoliubov transformation,
\begin{align}
\phi_{k_{x}^{a}}  &  =u\alpha+v\beta^{\dagger},\text{ }\phi_{k_{x}^{a}%
}^{\dagger}=u\alpha^{\dagger}+v\beta,\nonumber \\
\phi_{-k_{x}^{a}}^{\dagger}  &  =v\alpha+u\beta^{\dagger},\text{ }\phi
_{-k_{x}^{a}}=v\alpha^{\dagger}+u\beta,
\end{align}
where $u$ and $v$ satisfy the following conditions
\begin{align}
u^{2}  &  =\frac{1}{2}+\frac{2g+2h_{x}\cos k_{x}^{a}}{2E_{k^{a}}},\nonumber \\
v^{2}  &  =-\frac{1}{2}+\frac{2g+2h_{x}\cos k_{x}^{a}}{2E_{k^{a}}},\nonumber \\
uv  &  =-\frac{2h^{x}\cos k_{x}^{a}}{2E_{k^{a}}}.
\end{align}

Finally, the effective Hamiltonian of e-particle (or m-particle) is obtained
as
\begin{equation}
\mathcal{\hat{H}}_{\mathrm{eff}}^{a}=\sum_{k_{x}^{a}}E_{k^{a}}(\alpha
_{k_{x}^{a}}^{\dagger}\alpha_{k_{x}^{a}}+\beta_{k_{x}^{a}}^{\dagger}%
\beta_{k_{x}^{a}}).
\end{equation}
where the energy spectra are real as
\begin{equation}
E_{k^{a}}=2g\sqrt{1+\frac{2h_{x}\cos k_{x}^{a}}{g}}.
\end{equation}
When the term $h_{x}$ becomes larger, the energy gap for e/m-particle closes
and a quantum phase transition from $Z_{2}$ topological order to spin
polarized order occurs. In Fig.3(b), we plot the energy spectra of
e/m-particle for the (Hermitian) Wen-plaquette model with $h_{x}=0.2$,
$h_{z}=0.01$.

\begin{figure}[ptb]
\includegraphics[clip,width=0.42\textwidth]{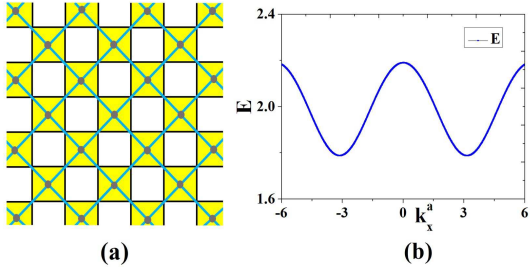}\caption{(Color online)
(a) The schematic diagram of effective lattice for m-particle. The grey dots
denote the lattice sites; (b) The energy spectra of boson for the (Hermitian)
Wen-plaquette model with $h_{x}=0.2$, $h_{z}=0.01$.}%
\end{figure}

On the other hand, with the help of perturbative method, the hopping term
$\hat{t}_{I^{f}J^{f}}^{f}$ for two types of f-particles on $I^{f}$-lattice
(lattice of links) corresponds to a one-step string operator $\hat{t}_{i}%
^{f}=h_{x}\sigma_{i}^{x}+h_{z}\sigma_{i}^{z}$, i.e.,
\begin{align}
\hat{t}_{i}^{f}  &  =h_{x}\sigma_{i}^{x}+h_{z}\sigma_{i}^{z}\nonumber \\
&  \rightarrow \hat{t}_{I^{f}J^{f}}^{f}=h_{x}(\phi_{f_{1},I^{f}}^{\dag}%
\phi_{f_{2},I^{f}+e_{x^{I^{f}}}^{I^{f}}}+\phi_{f_{1},I^{f}}^{\dag}\phi
_{f_{2},I^{f}+e_{x^{I^{f}}}^{I^{f}}}^{\dag}\nonumber \\
&  +\phi_{f_{1},I^{f}}^{\dag}\phi_{f_{2},I^{f}-e_{x^{I^{f}}}^{I^{f}}}%
+\phi_{f_{1},I^{f}}^{\dag}\phi_{f_{2},I^{f}-e_{x^{I^{f}}}^{I^{f}}}^{\dag
}+h.c.)\nonumber \\
&  +h_{z}(\phi_{f_{1},I^{f}}^{\dag}\phi_{f_{1},I^{f}+(e_{x^{I^{f}}}^{I^{f}%
}+e_{y^{I^{f}}}^{I^{f}})}+\phi_{f_{1},I^{f}}^{\dag}\phi_{f_{1},I^{f}%
+(e_{x^{I^{f}}}^{I^{f}}+e_{y^{I^{f}}}^{I^{f}})}^{\dag}\nonumber \\
&  +\phi_{f_{2},I^{f}}^{\dag}\phi_{f_{2},I^{f}+(e_{x^{I^{f}}}^{I^{f}%
}-e_{y^{I^{f}}}^{I^{f}})}+\phi_{f_{2},I^{f}}^{\dag}\phi_{f_{2},I^{f}%
+(e_{x^{I^{f}}}^{I^{f}}-e_{y^{I^{f}}}^{I^{f}})}^{\dagger}\nonumber \\
&  +h.c.),
\end{align}
where $\phi_{f_{1},I^{f}}^{\dag}$ and $\phi_{f_{2},I^{f}}^{\dag}$ are the
generation operator of the f-particles on two sub-links, respectively. Here,
$e_{x^{I^{f}}}^{I^{f}}$ and $e_{y^{I^{f}}}^{I^{f}}$ are the unit vector along
$x^{I^{f}}$ and $y^{I^{f}}$ direction on $I^{f}$-lattice, respectively, with
lattice constant to be $\frac{\sqrt{2}}{2}a_{0}$. See the illustration in Fig.4(a).

Then, the effective Hamiltonian of f-particle is obtained as
\begin{align}
\hat{H}_{\mathrm{eff}}^{f}=  &  \sum_{\langle I^{f},J^{f}\rangle}%
h_{x}\big[(\phi_{f_{1},I^{f}}^{\dag}\phi_{f_{2},I^{f}+e_{x^{I^{f}}}^{I^{f}}%
}+\phi_{f_{1},I^{f}}^{\dag}\phi_{f_{2},I^{f}+e_{x^{I^{f}}}^{I^{f}}}^{\dag
}\nonumber \\
&  +\phi_{f_{1},I^{f}}^{\dag}\phi_{f_{2},I^{f}-e_{x^{I^{f}}}^{I^{f}}}%
+\phi_{f_{1},I^{f}}^{\dag}\phi_{f_{2},I^{f}-e_{x^{I^{f}}}^{I^{f}}}^{\dag
})+h.c.\big]\nonumber \\
&  +\sum_{\langle I^{f},J^{f}\rangle}h_{z}\big[(\phi_{f_{1},I^{f}}^{\dag}%
\phi_{f_{1},I^{f}+(e_{x^{I^{f}}}^{I^{f}}+e_{y^{I^{f}}}^{I^{f}})}\nonumber \\
&  +\phi_{f_{1},I^{f}}^{\dag}\phi_{f_{1},I^{f}+(e_{x^{I^{f}}}^{I^{f}%
}+e_{y^{I^{f}}}^{I^{f}})}^{\dag}+\phi_{f_{2},I^{f}}^{\dag}\phi_{f_{2}%
,I^{f}+(e_{x^{I^{f}}}^{I^{f}}-e_{y^{I^{f}}}^{I^{f}})}\nonumber \\
&  +\phi_{f_{2},I^{f}}^{\dag}\phi_{f_{2},I^{f}+(e_{x^{I^{f}}}^{I^{f}%
}-e_{y^{I^{f}}}^{I^{f}})}^{\dagger}+h.c.\big]\nonumber \\
&  +4g\sum_{I^{f}}\phi_{f_{1},I^{f}}^{\dag}\phi_{f_{1},I^{f}}+4g\sum_{I^{f}%
}\phi_{f_{2},I^{f}}^{\dag}\phi_{f_{2},I^{f}}.
\end{align}
After Fourier transformation, $\hat{H}_{\mathrm{eff}}^{f}$ can be written as
\begin{equation}
\hat{H}_{\mathrm{eff}}^{f}=\left(
\begin{array}
[c]{cccc}%
\phi_{ok}^{\dagger} & \phi_{o-k} & \phi_{ek}^{\dagger} & \phi_{e-k}%
\end{array}
\right)  \hat{\mathcal{H}}_{\mathrm{eff}}^{f}\left(
\begin{array}
[c]{c}%
\phi_{ok}\\
\phi_{o-k}^{\dagger}\\
\phi_{ek}\\
\phi_{e-k}^{\dagger}%
\end{array}
\right)  ,
\end{equation}
where the effective Hamiltonian $\mathcal{\hat{H}}_{\mathrm{eff}}^{f}$ in
momentum space is obtained as%
\begin{equation}
\hat{\mathcal{H}}_{\mathrm{eff}}^{f}=\left(
\begin{array}
[c]{cccc}%
4g+A_{1} & iB_{1} & C & C\\
-iB_{1} & -4g-A_{1} & -C & -C\\
C & -C & 4g+A_{2} & iB_{2}\\
C & -C & -iB_{2} & -4g-A_{2}%
\end{array}
\right)  ,
\end{equation}
with $A_{1}=2h_{z}\cos(k_{x}^{f}+k_{y}^{f})$, $A_{2}=2h_{z}\cos(k_{x}%
^{f}-k_{y}^{f})$, $B_{1}=2h_{z}\sin(k_{x}^{f}+k_{y}^{f})$, $B_{2}=2h_{z}%
\sin(k_{x}^{f}-k_{y}^{f})$, $C=2h_{x}\cos k_{x}^{f}$. In Fig.4(b), we plot the
energy spectra of f-particle for the (Hermitian) Wen-plaquette model with
$h_{x}=0.2$, $h_{z}=0.01$.

\begin{figure}[ptb]
\includegraphics[clip,width=0.42\textwidth]{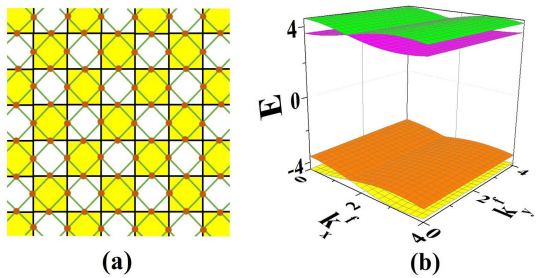}\caption{(Color online)
(a) The schematic diagram of effective lattice for f-particle. The red dots
denote the lattice sites; (b) The energy spectra of fermion for the
(Hermitian) Wen-plaquette model with $h_{x}=0.2$, $h_{z}=0.01$.}%
\end{figure}

\subsection{Degenerate ground states}

In this part, we review the degeneracy of ground states $\mathcal{D}$ for the
Wen-plaquette model. Under the periodic boundary condition (on a torus), the
degeneracy $\mathcal{D}$ is dependent on lattice number: $\mathcal{D}=4$ on
even-by-even ($e\ast e$) lattice, $\mathcal{D}=2$ on other cases (even-by-odd
($e\ast o$), odd-by-even ($o\ast e$) and odd-by-odd ($o\ast o$)
lattices)\cite{wen3,k1,k2,wen1,wen2,kou1,kou2}.

To classify the degeneracy of the ground states of $Z_{2}$ topological order,
we define topological closed string operators around torus.

{F}or the Wen-plaquette model on an $e\ast e$ lattice, there are four types of
topological closed string operators, $W_{v}(\mathcal{C}_{X}),$ $W_{v}%
(\mathcal{C}_{Y}),$ $W_{f}(\mathcal{C}_{X})$ and $W_{f}(\mathcal{C}_{Y}).$
Here $\mathcal{C}_{X}$ denotes a topological closed loop around the torus
along x-direction and $C_{Y}$ denotes a topological closed loop around the
torus along y-direction. The four types of topological closed string operators
are all Hermitian, i.e.,
\begin{align}
W_{v}(\mathcal{C}_{X})  &  =W_{v}^{\dagger}(\mathcal{C}_{X}),\text{ }%
W_{v}(\mathcal{C}_{Y})=W_{v}^{\dagger}(\mathcal{C}_{Y}),\nonumber \\
W_{f}(\mathcal{C}_{X})  &  =W_{f}^{\dagger}(\mathcal{C}_{X}),\text{ }%
W_{f}(\mathcal{C}_{Y})=W_{f}^{\dagger}(\mathcal{C}_{Y}).
\end{align}
Due to the commutation and anti-commutation relations between $W_{v}%
(\mathcal{C}_{A})$ and $W_{f}(\mathcal{C}_{B})$, $\left \{  W_{v}%
(\mathcal{C}_{X}),W_{f}(\mathcal{C}_{Y})\right \}  =0,$ $\left[  W_{v}%
(\mathcal{C}_{X}),W_{f}(\mathcal{C}_{X})\right]  =0,$ $\left \{  W_{v}%
(\mathcal{C}_{Y}),W_{f}(\mathcal{C}_{X})\right \}  =0,$ $\left[  W_{v}%
(\mathcal{C}_{Y}),W_{f}(\mathcal{C}_{Y})\right]  =0,$\ we may represent
$W_{v}(\mathcal{C}_{X}),$ $W_{v}(\mathcal{C}_{Y}),$ $W_{f}(\mathcal{C}_{X})$
and $W_{f}(\mathcal{C}_{Y})$ by pseudo-spin operators $\tau_{1}^{x},$
$\tau_{2}^{x},$ $\tau_{2}^{z}$ and $\tau_{1}^{z}$, respectively. The ground
states becomes the eigenstates of $\tau_{l}^{z}$ ($l=1,2$). Then one has four
degenerate ground states (denoted by $|m_{1},m_{2}\rangle=|m_{1}\rangle
\otimes|m_{2}\rangle$). For $m_{l}=0,$ we have $\tau_{l}^{z}|m_{l}%
\rangle=|m_{l}\rangle,$ and for $m_{l}=1$ we have $\tau_{l}^{z}|m_{l}%
\rangle=-|m_{l}\rangle.$ Physically, the topological degeneracy arises from
the presence or absence of $\pi$ flux of fermion through the hole. The values
of $m_{l}$ reflect the presence ($m_{l}=1$) or the absence ($m_{l}=0$) of the
$\pi$ flux in the hole. A ground state becomes the linear combination of the
four degenerate ground states $\left \vert 0\right \rangle =\displaystyle \sum
_{m_{1},m_{2}=0,1}\alpha_{m_{1},m_{2}} |m_{1},m_{2}\rangle$, where
$\alpha_{m_{1},m_{2}}$ is the weight of $|m_{1},m_{2}\rangle$.

Similarly we may use the closed string operators to describe the ground states
on $e\ast o$ lattice, $o\ast e$ lattice and $o\ast o$ lattice. For example,
for the Wen-plaquette model on an $e\ast o$ lattice, we can only define two
types of topological closed string operators, $W_{v}(\mathcal{C}_{X})$ and
$W_{f}(\mathcal{C}_{Y}).$ There is no topological closed string operator for
m-particles along y-direction with odd number of lattices. Due to the
anti-commutation relations between, $W_{v}(\mathcal{C}_{X})$ and
$W_{f}(\mathcal{C}_{Y})$, $\left \{  W_{v}(\mathcal{C}_{X}),W_{f}%
(\mathcal{C}_{Y})\right \}  =0,$\ we may represent $W_{v}(\mathcal{C}_{X})$ and
$W_{f}(\mathcal{C}_{Y})$ by pseudo-spin operators $\tau^{x}$ and $\tau^{z}$,
respectively. The ground states becomes the eigenstates of $\tau^{z}$. Then
one has two degenerate ground states (denoted by $|m\rangle$). For $m=0,$ we
have $\tau^{z}|m\rangle=|m\rangle,$ and for $m=1$ we have $\tau^{z}%
|m\rangle=-|m\rangle.$

It is known that the degenerate ground states of $Z_{2}$ topological orders
have same energy in thermodynamic limit. However, in a finite system,$\ $the
degeneracy of the ground states is (partially) removed due to tunneling
processes, of which a virtual quasi-particle moves around the torus before
annihilating with the other one\cite{k1,kou1,kou2}. For example, at first a
pair of the m-particle is created. One of the m-particles propagates all the
way around the torus and then annihilates with the other m-particle. Such a
process effectively adds a unit of the $\pi$-flux to the hole of the torus and
changes $m_{a}$ or $m_{b}$ by $1$. To characterize the low energy effective
physics of degenerate ground states, we derive the effective (Hermitian)
pseudo-spin Hamiltonian of the degenerate ground states.

\section{Non-Hermitian Wen-plaquette model with $\mathcal{PT}$-symmetry}

We introduce a non-Hermitian Wen-plaquette (toric-code) model with
$\mathcal{PT}$-symmetry, of which the Hamiltonian is
\begin{equation}
\hat{H}_{\mathrm{NTO}}=\hat{H}_{\mathrm{TO}}+\hat{H}_{\mathrm{PT}},
\end{equation}
where
\begin{equation}
\hat{H}_{\mathrm{TO}}=-{g\sum_{i}F_{i}},
\end{equation}
with ${F_{i}=\sigma_{i}^{x}\sigma_{i+\hat{e}_{x}}^{y}\sigma_{i+\hat{e}%
_{x}+\hat{e}_{y}}^{x}\sigma_{i+\hat{e}_{y}}^{y}}$ and
\begin{equation}
\hat{H}_{\mathrm{PT}}={\sum \limits_{i}}(h_{x}\sigma_{i}^{x}+ih_{z}\sigma
_{i}^{z}),
\end{equation}
$h_{x},h_{z}$ are two real parameters. $\sigma_{i}^{x,y,z}$ are Pauli matrices
on sites $i$.

We define parity operator $\mathcal{P}$ and time-reversal
operator$\  \mathcal{T}$, i.e., the time reversal operator $\mathcal{T}$ has
the function
\begin{equation}
\mathcal{T}i\mathcal{T=-}i,
\end{equation}
and the parity operator $\mathcal{P}$ has the function of rotating each spin
by $\pi$ about the $x$-axis
\begin{equation}
\mathcal{P}=%
%TCIMACRO{\dprod \limits_{j=1}^{N}}%
%BeginExpansion
{\displaystyle \prod \limits_{j=1}^{N}}
%EndExpansion
\sigma_{j}^{x}.
\end{equation}
As a result, the Hamiltonian $\hat{H}_{\mathrm{NTO}}$ is $\mathcal{PT}$
invariant
\begin{equation}
\left[  \mathcal{P},\hat{H}_{\mathrm{NTO}}\right]  \neq0\text{ and }\left[
\mathcal{T},\hat{H}_{\mathrm{NTO}}\right]  \neq0,
\end{equation}
but
\[
\left[  \mathcal{PT},\text{ }\hat{H}_{\mathrm{NTO}}\right]  =0.
\]

For the case of $g\gg h_{x},h_{z}$, $\hat{H}_{\mathrm{NTO}}\rightarrow \hat
{H}_{\mathrm{TO}},$ we have a $Z_{2}$ topological ordered state. In this
limit, the elementary excitations (m/e-particles) are defined as ${F_{i}=-1}$
at odd/even sub-plaquette. The bound states of an e-particle and a m-particle
obey fermionic statistic. For the case of $g\ll h_{x},h_{z}$, $\hat
{H}_{\mathrm{NTO}}\rightarrow \hat{H}_{\mathrm{PT}}$, we have a typical
non-Hermitian Hamiltonian with spontaneous $\mathcal{PT}$-symmetry breaking.

In this paper, we focus on the case of $g\gg h_{x},h_{z}$ and treat the
non-Hermitian term $\hat{H}_{\mathrm{PT}}$ as a perturbation. Now, the ground
state becomes a non-Hermitian $Z_{2}$ topological ordered state.

\section{Biorthogonal description for non-Hermitian dynamic strings}

\begin{figure}[ptb]
\includegraphics[clip,width=0.42\textwidth]{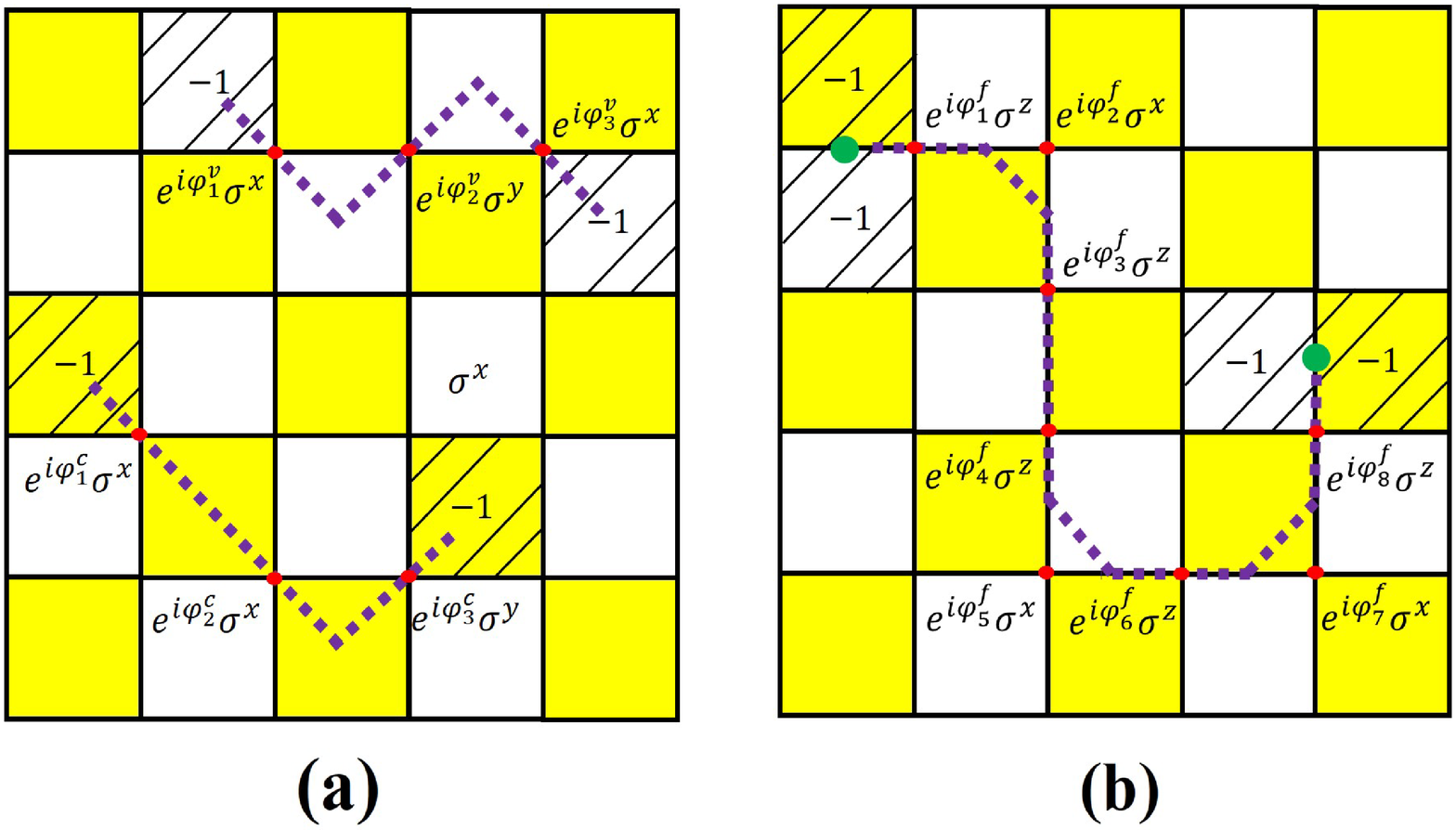}\caption{(Color online)
(a) The schematic diagram of dynamic strings for e/m-particles under the
perturbation; (b) The schematic diagram of dynamic strings for f-particle
under the perturbation.}%
\end{figure}

\subsection{Dynamic strings for the non-Hermitian $Z_{2}$ topological order}

To characterize the quantum properties of the non-Hermitian topological order
for $\hat{H}_{\mathrm{NTO}}$, we introduce three types of strings along a
$N$-step loop $\mathcal{C}_{N}$: the operation strings $W_{a}(\mathcal{C}%
_{N})$, the phase strings $P_{a}(\mathcal{C}_{N})$, and the dynamic strings
$D_{a}(\mathcal{C}_{N})$, respectively. The indices $a=v,$ $c,$ $f$ correspond
to three types of quasi-particles.

An operation string is defined as the product of spin operators,
\begin{equation}
W_{a}(\mathcal{C}_{N})=\prod_{i\in C}\sigma_{i}^{\alpha_{i}},
\end{equation}
where $\prod_{i\in \mathcal{C}}$ is over all the sites on the string along a
loop $\mathcal{C}$. $\sigma_{i}^{\alpha_{i}}$ is $\alpha_{i}$-type Pauli
matrix on site $i$. A phase string is introduced as the product of phases
along a loop $\mathcal{C},$
\begin{equation}
P_{a}(\mathcal{C}_{N})=\prod_{i\in C}e^{i\varphi_{i}^{a}},
\end{equation}
where $\varphi_{i}^{a}$ is the phase of the quantum states at step $i$ for
$a$-type excitation. The dynamic string $D_{a}(\mathcal{C}_{N})$ is introduced
to characterize the hopping property, i.e.,
\begin{equation}
D_{a}(\mathcal{C}_{N})=\prod_{i\in \mathcal{C}}\frac{\hat{t}_{i}^{a}%
}{\left \vert \hat{t}_{i}^{a}\right \vert },
\end{equation}
where
\begin{equation}
\frac{\hat{t}_{i}^{a}}{\left \vert \hat{t}_{i}^{a}\right \vert }=e^{i\varphi
_{i}^{a}}\sigma_{i}^{\alpha_{i}}.
\end{equation}
For a given topological excitation ($a=v,$ $c,$ or $f$), a dynamic string is a
combination of an operation string and a phase string, according to the
relationship of the three types of strings, i.e.,
\begin{equation}
D_{a}(\mathcal{C}_{N})=W_{a}(\mathcal{C}_{N})P_{a}(\mathcal{C}_{N}).
\end{equation}

The phase strings $P_{a}(\mathcal{C}_{N})$ and the dynamic strings
$D_{a}(\mathcal{C}_{N})$ can be Hermitian or non-Hermitian. We then define
Hermitian/non-Hermitian dynamic string $D_{a}(\mathcal{C}_{N})$ by verifying
its conjugation. For the case of
\begin{equation}
D_{a}(\mathcal{C}_{N})=D_{a}^{\dagger}(\mathcal{C}_{N}),
\end{equation}
a dynamic string is Hermitian; For the case of
\begin{equation}
D_{a}(\mathcal{C}_{N})\neq D_{a}^{\dagger}(\mathcal{C}_{N}),
\end{equation}
a dynamic string is non-Hermitian.

\subsection{Biorthogonal set for the quantum string states}

To characterize the non-Hermitian dynamic string $D_{a}(\mathcal{C}_{N})\neq
D_{a}^{\dagger}(\mathcal{C}_{N})$, we introduce the \emph{biorthogonal set}
for the quantum string states of the non-Hermitian $Z_{2}$ topological order\cite{ma}.

In general, for non-Hermitian dynamic strings, we define left/right
eigenstates defined as
\begin{equation}
D_{a}(\mathcal{C}_{N})|{\Psi}^{\mathrm{R}}\rangle=d_{a}(\mathcal{C}_{N}%
)|{\Psi}^{\mathrm{R}}\rangle,
\end{equation}
and
\begin{equation}
D_{a}^{\dagger}(\mathcal{C}_{N})|{\Psi}^{\mathrm{L}}\rangle=d_{a}^{\ast
}(\mathcal{C}_{N})|{\Psi}^{\mathrm{L}}\rangle,
\end{equation}
where $d_{a}(\mathcal{C}_{N})$, $d_{a}^{\ast}(\mathcal{C}_{N})$ are the
corresponding eigenvalues. The eigenstates can be bi-orthogonalized as
\begin{equation}
|\widetilde{{\Psi}}_{n}^{\mathrm{R}}\rangle=|{\Psi}_{n}^{\mathrm{R}}%
\rangle/\sqrt{\langle{\Psi}_{n}^{\mathrm{L}}|{\Psi}_{n}^{\mathrm{R}}\rangle},
\end{equation}
and
\begin{equation}
|\widetilde{{\Psi}}_{n}^{\mathrm{L}}\rangle=|{\Psi}_{n}^{\mathrm{L}}%
\rangle/\sqrt{\langle{\Psi}_{n}^{\mathrm{R}}|{\Psi}_{n}^{\mathrm{L}}\rangle},
\end{equation}
which then gives
\begin{equation}
\langle{\widetilde{\Psi}}_{n}^{\mathrm{L}}|\widetilde{{\Psi}}_{m}^{\mathrm{R}%
}\rangle=\delta_{nm},
\end{equation}
and
\begin{equation}
\sum_{n}|{\widetilde{\Psi}}_{m}^{\mathrm{R}}\rangle \langle{\widetilde{\Psi}%
}_{n}^{\mathrm{L}}|=1.
\end{equation}
As a result, we have
\begin{equation}
D_{a}^{\dagger}(\mathcal{C}_{N})|\widetilde{\Psi}_{n}^{\mathrm{R}}%
\rangle=d_{a}(\mathcal{C}_{N})|\widetilde{\Psi}_{n}^{\mathrm{R}}\rangle.
\end{equation}

In particular, a dynamic string is $\mathcal{PT}$-symmetric non-Hermitian when
it satisfies the following conditions,
\begin{equation}
\left[  \mathcal{P},D_{a}(\mathcal{C}_{N})\right]  \neq0,~~\left[
\mathcal{T},D_{a}(\mathcal{C}_{N})\right]  \neq0,
\end{equation}
but
\begin{equation}
\left[  \mathcal{PT},\text{ }D_{a}(\mathcal{C}_{N})\right]  =0.
\end{equation}

\subsection{Dynamic strings for non-Hermitian Wen-plaquette model}

For the non-Hermitian topological order described by $\hat{H}_{\mathrm{NTO}},$
we show the detailed definitions for three types of strings.

To create an e-particle/m-particle, one define the operation string that
connects the nearest neighboring even (odd) sub-plaquettes
\begin{equation}
W_{c/v}(\mathcal{C}_{N})=\prod_{i\in \mathcal{C}}\sigma_{i}^{\alpha_{i}}.
\end{equation}
The product $\prod_{i\in \mathcal{C}}$ is over all the sites on the string
along a loop $\mathcal{C}$ connecting even-plaquettes (or odd-plaquettes),
$\alpha_{i}=x$ if $i$ is odd and $\alpha_{i}=y$ if $i$ is even. Under the
perturbation $h_{x}%
%TCIMACRO{\dsum \limits_{i}}%
%BeginExpansion
{\displaystyle \sum \limits_{i}}
%EndExpansion
\sigma_{i}^{x},$ the e-particle/m-particle begins to hop along $\hat{e}%
_{x}-\hat{e}_{y}$ direction. As a result, we have $\hat{t}_{i}^{v/c}%
=h_{x}\sigma_{i}^{x}.$ The phase string and dynamic string for
e-particle/m-particle are obtained as
\begin{align}
P_{c}(\mathcal{C}_{N})  &  =P_{v}(\mathcal{C}_{N})=1,\nonumber \\
D_{c}(\mathcal{C}_{N})  &  =D_{v}(\mathcal{C}_{N})=\prod_{i\in \mathcal{C}%
}\sigma_{i}^{x},
\end{align}
respectively.

The operation string for f-particles is defined as
\begin{equation}
W_{f}(\mathcal{C}_{N})=\prod_{i\in \mathcal{C}}\sigma_{i}^{\alpha_{i}},
\end{equation}
where $\alpha_{i}=z$ if the string does not turn at site $i$, $\alpha_{i}=x$
if the turn forms a upper-left or lower-right corner, $\alpha_{i}=y$ if the
turn forms a lower-left or upper-right corner. Under a perturbation
$ih_{z}\sum \limits_{i}\sigma_{i}^{z}$, f-particles begin to move. Now, we have
$\hat{t}_{i}^{f}=ih_{z}\sigma_{i}^{z}$ (without considering the term $h_{x}%
%TCIMACRO{\dsum \limits_{i}}%
%BeginExpansion
{\displaystyle \sum \limits_{i}}
%EndExpansion
\sigma_{i}^{x}$). The phase string and dynamic string for fermion are obtained
as
\begin{align}
P_{f}(\mathcal{C}_{N})  &  =\prod_{i\in \mathcal{C}}e^{i\varphi_{i}^{a}%
},\nonumber \\
D_{f}(\mathcal{C}_{N})  &  =\prod_{i\in \mathcal{C}}e^{i\varphi_{i}^{a}}%
\sigma_{i}^{z}.
\end{align}
with $\varphi_{i}^{a}=\frac{\pi}{2}$. For the case of even number $N$, the
phase string is $P_{f}(\mathcal{C}_{N})=\pm1$ and the dynamic string is
\begin{equation}
D_{f}(\mathcal{C}_{N})=\pm \prod_{i\in \mathcal{C}}\sigma_{i}^{z}=\pm
W_{f}(\mathcal{C}_{N}).
\end{equation}
The phase string and dynamic string for fermions are Hermitian($D_{f}^{\dag
}(\mathcal{C}_{N})=D_{f}(\mathcal{C}_{N})$). For the case of odd number $N$,
the phase string is $P_{f}(\mathcal{C}_{N})=e^{\pm i\frac{\pi}{2}}$ and the
dynamic string is
\begin{equation}
D_{f}(\mathcal{C}_{N})=e^{\pm i\frac{\pi}{2}}\prod_{i\in \mathcal{C}}\sigma
_{i}^{z}=e^{\pm i\frac{\pi}{2}}W_{f}(\mathcal{C}_{N}),
\end{equation}
that is $\mathcal{PT}$-symmetric non-Hermitian according to
\begin{equation}
\left[  \mathcal{P},D_{f}(\mathcal{C}_{N})\right]  \neq0,\text{ }\left[
\mathcal{T},D_{f}(\mathcal{C}_{N})\right]  \neq0,
\end{equation}
and
\begin{equation}
\left[  \mathcal{PT},\text{ }D_{f}(\mathcal{C}_{N})\right]  =0.
\end{equation}

\section{Quasi-particles}

The open string creates two point-like objects that correspond to different
types of quasi-particles (topological excitations) at its ends. In this part,
we use the perturbative method to describe the effective Hamiltonian of
quasi-particles. These effective Hamiltonians describe a hard-core
boson/fermion system.

\subsection{Energy spectra for e/m-particle}

By perturbative method, each hopping term $\hat{t}_{I^{a}J^{a}}^{a}$ for
e/m-particle on $I^{a}$-lattice ($a=v$ on lattice of odd sub-plaquettes and
$a=c$ on lattice of even sub-plaquettes) corresponds to a one-step Hermitian
dynamic string
\begin{align}
\hat{t}_{i}^{a}  &  =h_{x}\sigma_{i}^{x}\rightarrow \hat{t}_{I^{a}J^{a}}%
^{a}=h_{x}(\phi_{a,I^{a}}^{\dag}\phi_{a,I^{a}+e_{x^{I^{a}}}^{I^{a}}%
}\nonumber \\
&  +\phi_{a,I^{a}}^{\dag}\phi_{a,I^{a}+e_{x^{I^{a}}}^{I^{a}}}^{\dag}+h.c.),
\end{align}
where $\phi_{a,I^{a}}^{\dag}$ is the generation operator of the e/m-particle.
Here, $e_{x^{I^{a}}}^{I^{a}}$ is the unit vector along $x^{I^{a}}$ direction
on $I^{a}$-lattice, of which the lattice constant is $\sqrt{2}a_{0}$ ($a_{0}$
is lattice constant of the original square lattice).

The perturbative effective Hamiltonian of boson $\hat{H}_{\mathrm{eff}}^{a}$
($a=v$ denote vortex excitation, $a=c$ denote charge excitation) on
$L_{x}\times L_{y}$ square lattice can be expressed as
\begin{align}
\hat{H}_{\mathrm{eff}}^{a}=  &  \sum_{\left \langle I^{a},J^{a}\right \rangle
}h_{x}(\phi_{a,I}^{\dag}\phi_{a,I^{a}+e_{x^{I^{a}}}^{I^{a}}}+\phi_{a,I}^{\dag
}\phi_{a,I^{a}+e_{x^{I^{a}}}^{I^{a}}}^{\dag}+h.c.)\nonumber \\
&  +2g\sum_{I^{a}}\phi_{a,I^{a}}^{\dag}\phi_{a,I^{a}}.
\end{align}
The effective Hamiltonian $\mathcal{\hat{H}}_{\mathrm{eff}}^{a}$ in momentum
space of e/m-particle can be obtained as $\mathcal{\hat{H}}_{\mathrm{eff}}%
^{a}=\sum_{k_{x}^{a}}E_{k^{a}}(\alpha_{k_{x}^{a}}^{\dagger}\alpha_{k_{x}^{a}%
}+\beta_{k_{x}^{a}}^{\dagger}\beta_{k_{x}^{a}})$ where the energy spectra can
be expressed as
\begin{equation}
E_{k^{a}}=\sqrt{(2g+2h_{x}\cos k_{x}^{a})^{2}-(2h^{x}\cos k_{x}^{a})^{2}}.
\end{equation}
The result is same as the Hermitian case.

\subsection{Non-Hermitian lattice model and complex spectra for f-particle}

In the perturbative method, the hopping term $\hat{t}_{I^{f}J^{f}}^{f}$ for
two types of f-particle on $I^{f}$-lattice (lattice of links) corresponds to a
one-step non-Hermitian dynamic string $\hat{t}_{i}^{f}=h_{x}\sigma_{i}%
^{x}+ih_{z}\sigma_{i}^{z}$, i.e.,
\begin{align}
\hat{t}_{i}^{f}  &  =h_{x}\sigma_{i}^{x}+ih_{z}\sigma_{i}^{z}\nonumber \\
&  \rightarrow \hat{t}_{I^{f}J^{f}}^{f}=h_{x}(\phi_{f_{1},I^{f}}^{\dag}%
\phi_{f_{2},I^{f}+e_{x^{I^{f}}}^{I^{f}}}+\phi_{f_{1},I^{f}}^{\dag}\phi
_{f_{2},I^{f}+e_{x^{I^{f}}}^{I^{f}}}^{\dag}\nonumber \\
&  +\phi_{f_{1},I^{f}}^{\dag}\phi_{f_{2},I^{f}-e_{x^{I^{f}}}^{I^{f}}}%
+\phi_{f_{1},I^{f}}^{\dag}\phi_{f_{2},I^{f}-e_{x^{I^{f}}}^{I^{f}}}^{\dag
}+h.c.)\nonumber \\
&  +ih_{z}(\phi_{f_{1},I^{f}}^{\dag}\phi_{f_{1},I^{f}+(e_{x^{I^{f}}}^{I^{f}%
}+e_{y^{I^{f}}}^{I^{f}})}+\phi_{f_{1},I^{f}}^{\dag}\phi_{f_{1},I^{f}%
+(e_{x^{I^{f}}}^{I^{f}}+e_{y^{I^{f}}}^{I^{f}})}^{\dag}\nonumber \\
&  +\phi_{f_{2},I^{f}}^{\dag}\phi_{f_{2},I^{f}+(e_{x^{I^{f}}}^{I^{f}%
}-e_{y^{I^{f}}}^{I^{f}})}+\phi_{f_{2},I^{f}}^{\dag}\phi_{f_{2},I^{f}%
+(e_{x^{I^{f}}}^{I^{f}}-e_{y^{I^{f}}}^{I^{f}})}^{\dagger}\nonumber \\
&  +h.c.),
\end{align}
where $\phi_{f_{1},I^{f}}^{\dag}$ and $\phi_{f_{2},I^{f}}^{\dag}$ are the
generation operator of the fermionic quasi-particles on two sub-links,
respectively. Here, $e_{x^{I^{f}}}^{I^{f}}$ and $e_{y^{I^{f}}}^{I^{f}}$ are
the unit vector along $x^{I^{f}}$ and $y^{I^{f}}$ direction on $I^{f}%
$-lattice, respectively, with lattice constant to be $\frac{\sqrt{2}}{2}a_{0}%
$. In particular, the hopping term $\hat{t}_{I^{f}J^{f}}^{f}=h_{x}\sigma
_{i}^{x}+ih_{z}\sigma_{i}^{z}$ corresponding to a one-step non-Hermitian
dynamic string becomes non-Hermitian, i.e., $\hat{t}_{I^{f}J^{f}}^{f}\neq
(\hat{t}_{I^{f}J^{f}}^{f})^{\dagger}$.

Consequently, the perturbative effective Hamiltonian of f-particle becomes
non-Hermitian that is obtained as
\begin{align}
\hat{H}_{\mathrm{eff}}^{f}=  &  \sum_{\langle I^{f},J^{f}\rangle}%
h_{x}\big[(\phi_{f_{1},I^{f}}^{\dag}\phi_{f_{2},I^{f}+e_{x^{I^{f}}}^{I^{f}}%
}+\phi_{f_{1},I^{f}}^{\dag}\phi_{f_{2},I^{f}+e_{x^{I^{f}}}^{I^{f}}}^{\dag
}\nonumber \\
&  +\phi_{f_{1},I^{f}}^{\dag}\phi_{f_{2},I^{f}-e_{x^{I^{f}}}^{I^{f}}}%
+\phi_{f_{1},I^{f}}^{\dag}\phi_{f_{2},I^{f}-e_{x^{I^{f}}}^{I^{f}}}^{\dag
})+h.c.\big]\nonumber \\
&  +\sum_{\langle I^{f},J^{f}\rangle}ih_{z}\big[(\phi_{f_{1},I^{f}}^{\dag}%
\phi_{f_{1},I^{f}+(e_{x^{I^{f}}}^{I^{f}}+e_{y^{I^{f}}}^{I^{f}})}\nonumber \\
&  +\phi_{f_{1},I^{f}}^{\dag}\phi_{f_{1},I^{f}+(e_{x^{I^{f}}}^{I^{f}%
}+e_{y^{I^{f}}}^{I^{f}})}^{\dag}+\phi_{f_{2},I^{f}}^{\dag}\phi_{f_{2}%
,I^{f}+(e_{x^{I^{f}}}^{I^{f}}-e_{y^{I^{f}}}^{I^{f}})}\nonumber \\
&  +\phi_{f_{2},I^{f}}^{\dag}\phi_{f_{2},I^{f}+(e_{x^{I^{f}}}^{I^{f}%
}-e_{y^{I^{f}}}^{I^{f}})}^{\dagger}+h.c.\big]\nonumber \\
&  +4g\sum_{I^{f}}\phi_{f_{1},I^{f}}^{\dag}\phi_{f_{1},I^{f}}+4g\sum_{I^{f}%
}\phi_{f_{2},I^{f}}^{\dag}\phi_{f_{2},I^{f}}.
\end{align}
After Fourier transformation, the effective Hamiltonian $\hat{H}%
_{\mathrm{eff}}^{f}$ can be written as
\begin{equation}
\hat{H}_{\mathrm{eff}}^{f}=\left(
\begin{array}
[c]{cccc}%
\phi_{ok}^{\dagger} & \phi_{o-k} & \phi_{ek}^{\dagger} & \phi_{e-k}%
\end{array}
\right)  \hat{\mathcal{H}}_{\mathrm{eff}}^{f}\left(
\begin{array}
[c]{c}%
\phi_{ok}\\
\phi_{o-k}^{\dagger}\\
\phi_{ek}\\
\phi_{e-k}^{\dagger}%
\end{array}
\right)  ,
\end{equation}
where the effective Hamiltonian $\mathcal{\hat{H}}_{\mathrm{eff}}^{f}$ in
momentum space is obtained as%
\begin{equation}
\hat{\mathcal{H}}_{\mathrm{eff}}^{f}=\left(
\begin{array}
[c]{cccc}%
4g+iA_{1} & -B_{1} & C & C\\
B_{1} & -4g-iA_{1} & -C & -C\\
C & -C & 4g+iA_{2} & -B_{2}\\
C & -C & B_{2} & -4g-iA_{2}%
\end{array}
\right)
\end{equation}
with
\begin{align}
A_{1}  &  =2h_{z}\cos(k_{x}^{f}+k_{y}^{f}),\text{ }A_{2}=2h_{z}\cos(k_{x}%
^{f}-k_{y}^{f}),\nonumber \\
B_{1}  &  =2h_{z}\sin(k_{x}^{f}+k_{y}^{f}),\text{ }B_{2}=2h_{z}\sin(k_{x}%
^{f}-k_{y}^{f}),\nonumber \\
C  &  =2h_{x}\cos k_{x}^{f}.
\end{align}
As a result, the Hamiltonian $\mathcal{\hat{H}}_{\mathrm{eff}}^{f}$ can be
rewritten as%
\begin{align}
\mathcal{\hat{H}}_{\mathrm{eff}}^{f}  &  =4g(\mathbb{I\otimes}{\tau^{z}%
)}+\frac{1}{2}iA_{1}(\mathbb{I\otimes}{\tau^{z}+\sigma^{z}\otimes \tau^{z}%
)}\nonumber \\
&  +\frac{1}{2}iA_{2}(\mathbb{I\otimes}{\tau^{z}-{\sigma^{z}}\otimes \tau^{z}%
)}\nonumber \\
&  +\frac{1}{2}B_{1}\mathbf{(}-\mathbb{I\otimes}i{\tau^{y}-{\sigma^{z}}\otimes
i\tau^{y})}\nonumber \\
&  +\frac{1}{2}B_{2}\mathbf{(}-\mathbb{I\otimes}i{\tau^{y}+{\sigma^{z}}\otimes
i\tau^{y})}\nonumber \\
&  +C({{\sigma^{x}}\otimes \tau^{z}-{\sigma^{y}}\otimes \tau^{y})},
\end{align}
where ${\sigma^{i},}$ ${\tau^{i}}$ $(i=x,y,z)$ are Pauli matrices,
$\mathbb{I}$ is a $2\times2$ identity matrix.

\begin{figure}[ptb]
\includegraphics[clip,width=0.42\textwidth]{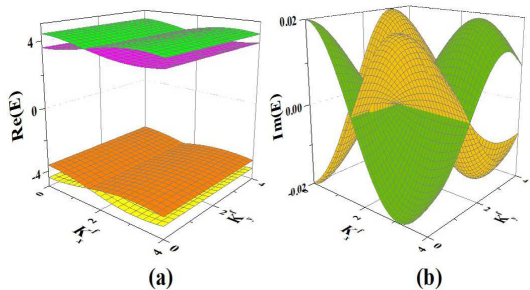}\caption{(Color online)
(a) The real part of energy spectrum of f-particle for the non-Hermitian
Wen-plaquette model with $h_{x}=0.2$, $h_{z}=0.01;$ (b) The imaginary part of
energy spectrum of fermion for the non-Hermitian Wen-plaquette model with
$h_{x}=0.2$, $h_{z}=0.01.$}%
\end{figure}

By diagonalizing this Hamiltonian, we can obtain the energy spectra for
f-particles that are tedious. Therefore, we focus on some special points in
momentum space. For example, at $k_{x}^{f}=\pm \frac{\pi}{2},$ $k_{y}^{f}=0$,
the effective Hamiltonian is reduced into
\begin{equation}
\hat{H}_{\mathrm{eff}}^{f}=4g(\mathbb{I\otimes}{\tau^{z})}\pm2ih_{z}%
{(}\mathbb{I\otimes}{\tau^{y}).}%
\end{equation}
The corresponding energy spectrum turns into%
\begin{equation}
E_{\mathrm{eff}}^{f}=\pm \sqrt{\left(  4g\right)  ^{2}-\left(  2h_{z}\right)
^{2}}.
\end{equation}
A $\mathcal{PT}$-symmetry-breaking transition occurs at the exceptional points
$E_{\mathrm{eff}}^{f}=0$ that leads to the following relation $2g=h_{z}.$

In Fig.6, we plot the real part and imaginary part of energy spectra of
f-particle for the non-Hermitian Wen-plaquette model with $h_{x}=0.2$,
$h_{z}=0.01$. One can see that the energy spectra of f-particle are complex
due to the non-Hermitian term, which is different from the Hermitian case.

\section{Anomalous topological degeneracy}

It was known that the ground state for $\hat{H}_{\mathrm{TO}}$ is a $Z_{2}$
topological order, of which the ground states have topological degeneracy,
i.e., different topologically degenerate ground states are classified by
different topological closed strings $W_{a}(\mathcal{C}_{N}%
^{\mathrm{close,topo}})$ that are all Hermitian, $W_{a}(\mathcal{C}%
_{N}^{\mathrm{close,topo}})=W_{a}^{\dagger}(\mathcal{C}_{N}%
^{\mathrm{close,topo}})$. Under the periodic boundary condition (on a torus)
for $\hat{H}_{\mathrm{TO}}$, the degeneracy is dependent on lattice number,
i.e., $4$ on a even-by-even ($e\ast e$) lattice, $2$ on even-by-odd ($e\ast
o$), odd-by-even ($o\ast e$) and odd-by-odd ($o\ast o$) lattices. Thus we get
a four-level (or two-level) system which can be mapped onto an effective
pseudo-spin model for topologically degenerate ground states\cite{kou1,kou2}.

After considering the non-Hermitian perturbation $\hat{H}_{\mathrm{PT}},$ the
ground states are classified by different topological closed dynamic strings
around the torus $D_{a}(\mathcal{C}_{N}^{\mathrm{close,topo}})$ that may be
either Hermitian or non-Hermitian. Under the periodic boundary condition (on a
torus) for $\hat{H}_{\mathrm{NTO}}$, the degeneracy is also dependent on
lattice number, i.e., $4$ on a even-by-even ($e\ast e$) lattice, $2$ on
even-by-odd ($e\ast o$), odd-by-even ($o\ast e$) and odd-by-odd ($o\ast o$)
lattices. The effective model for the topologically degenerate ground states
can also be mapped onto an effective pseudo-spin model.\ After considering the
quantum tunneling processes that change one ground state to another, the
transfer matrices between different topologically degenerate ground states are
characterized by the expectation values of topological closed dynamic strings
$D_{a}(\mathcal{C}_{N}^{\mathrm{close,topo}})$. Consequently, the effective
pseudo-spin model for topologically degenerate ground states may be either
Hermitian or non-Hermitian. In the following parts, we will derive the
effective (Hermitian or non-Hermitian) pseudo-spin model for topologically
degenerate ground states by calculating the expectation values for different
topological closed dynamic strings around the torus $D_{a}(\mathcal{C}%
_{N}^{\mathrm{close,topo}})$.

For the case of $e\ast e$ lattice, the topological closed dynamic strings
$D_{a}(\mathcal{C}_{N}^{\mathrm{close}})$ ($a=v,c$ or $f$) that correspond to
different quantum tunneling processes are all Hermitian, the effective
pseudo-spin model $\mathcal{\hat{H}}_{\mathrm{eff}}^{e\ast e}$ for
topologically degenerate ground states is also Hermitian. Therefore, in this
part we focus on the case of non-Hermitian Wen-plaquette model $\hat
{H}_{\mathrm{NTO}}$ on $o\ast e$ or $e\ast o,$ $o\ast o$ lattice.

\subsection{Non-Hermitian higher Order perturbation theory}

Firstly, a non-Hermitian higher order perturbation theory is developed to
calculate the energy splitting of the two degenerate ground states of
non-Hermitian Wen-plaquette model on $o\ast e$ or $e\ast o,$ $o\ast o$
lattice. For $m_{l}=0,$ we have $\tau_{l}^{z}|m_{l}\rangle=|m_{l}\rangle,$ and
for $m_{l}=1$ we have $\tau_{l}^{z}|m_{l}\rangle=-|m_{l}\rangle.$ We denote
$|0\rangle$ and $|1\rangle$ by $\mid \uparrow \rangle$ and $\left \vert
\downarrow \right \rangle ,$ respectively.

$\mid \uparrow \rangle_{R}$ and $\mid \uparrow \rangle_{L}$ is one of the ground
state of $\hat{H}_{0}$ and $\hat{H}_{0}^{\dagger}$ under biorthogonal set,
$\mid \downarrow \rangle_{R}$ and $\mid \downarrow \rangle_{L}$ is another ground
state of $\hat{H}_{0}$ and $\hat{H}_{0}^{\dagger}$ under biorthogonal set.
Under perturbation $\hat{H}_{I}$, $\hat{H}_{0}$ turns into $\hat{H}$. Now, the
quantum states of $\mid \uparrow \rangle_{R}$ and $\mid \downarrow \rangle_{R}$
turn into $\left \vert \Phi_{\uparrow}\right \rangle _{R}$ and $\left \vert
\Phi_{\downarrow}\right \rangle _{R}$ that can be constructed as following%
\begin{align}
\left \vert \Phi_{\uparrow}\right \rangle _{R}  &  =\frac{\hat{U}_{I}%
(0,-\infty)\left \vert \uparrow \right \rangle _{R}}{{}_{L}\left \langle
\uparrow \right \vert \hat{U}_{I}(0,-\infty)\left \vert \uparrow \right \rangle
_{R}},\label{zero}\\
\left \vert \Phi_{\downarrow}\right \rangle _{R}  &  =\frac{\hat{U}%
_{I}(0,-\infty)\left \vert \downarrow \right \rangle _{R}}{{}_{L}\left \langle
\downarrow \right \vert \hat{U}_{I}(0,-\infty)\left \vert \downarrow \right \rangle
_{R}}, \label{one}%
\end{align}
which is the eigenstate of Hamiltonian $\hat{H}$.

The transformation operator $\hat{U_{I}}(0,-\infty)$ can be written as
\begin{equation}
\hat{U}_{I}(0,-\infty)=\mathrm{T}\exp(-i\int_{-\infty}^{0}\hat{H}_{I}(t)dt)
\end{equation}
Here T denotes a time order and $\hat{H}_{I}(t)=e^{i\frac{\hat{H}_{0}}{\hbar
}t}\hat{H}_{I}e^{-i\frac{\hat{H}_{0}}{\hbar}t}$. Then the transformation
operator $\hat{U}_{I}(0,-\infty)\left \vert \uparrow \right \rangle _{R}$ in
Eq.\ref{zero} can be rewritten as
\begin{equation}
\hat{U}_{I}(0,-\infty)\left \vert \uparrow \right \rangle _{R}=\sum
\limits_{j=0}^{\infty}\hat{U}_{I}^{(j)}(0,-\infty)\left \vert \uparrow
\right \rangle _{R},\label{U0}%
\end{equation}
in which
\[
\hat{U}_{I}^{(0)}(0,-\infty)\left \vert \uparrow \right \rangle _{R}=\left \vert
\uparrow \right \rangle _{R}%
\]
and%
\begin{align}
\hat{U}_{I}^{(1)}(0,-\infty)\left \vert \uparrow \right \rangle _{R}= &
-\frac{i}{\hbar}\int_{-\infty}^{0}\hat{H}_{I}(t)dt\left \vert \uparrow
\right \rangle _{R}\nonumber \\
= &  \frac{1}{E_{0\uparrow}-\hat{H}_{0}}\hat{H}_{I}\left \vert \uparrow
\right \rangle _{R},
\end{align}
and
\begin{align}
\hat{U}_{I}^{(2)}(0,-\infty)\left \vert \uparrow \right \rangle _{R}= &
-\frac{i}{\hbar}\int_{-\infty}^{0}\hat{H}_{I}(t)\hat{U}_{I}^{(1)}%
(0,-\infty)dt\left \vert \uparrow \right \rangle _{R}\nonumber \\
= &  \frac{1}{E_{0\uparrow}-\hat{H}}_{0}\hat{H}_{I}\frac{1}{E_{0\uparrow}%
-\hat{H}_{0}}\hat{H}_{I}\left \vert \uparrow \right \rangle _{R},
\end{align}
and
\begin{equation}
\hat{U}_{I}^{(j\neq0)}(0,-\infty)\left \vert \uparrow \right \rangle _{R}%
=(\frac{1}{E_{0\uparrow}-\hat{H}_{0}}\hat{H_{I}})^{j}\left \vert \uparrow
\right \rangle _{R}%
\end{equation}
where $E_{\uparrow}$ is the eigenvalue of $\left \vert \Phi_{\uparrow
}\right \rangle _{R}$ for Hamiltonian $\hat{H}$, and $E_{0\uparrow}$ is the
eigenvalue of $\left \vert \uparrow \right \rangle _{R}$ for Hamiltonian $\hat
{H}_{0}$. Therefore, we have
\begin{equation}
\hat{U}_{I}(0,-\infty)\left \vert \uparrow \right \rangle _{R}=\left \vert
\uparrow \right \rangle _{R}+\sum \limits_{j=1}^{\infty}(\frac{1}{E_{0\uparrow
}-\hat{H}_{0}}\hat{H_{I}})^{j}\left \vert \uparrow \right \rangle _{R}.
\end{equation}
Similarity,
\begin{equation}
\hat{U}_{I}(0,-\infty)\left \vert \downarrow \right \rangle _{R}=\left \vert
\downarrow \right \rangle _{R}+\sum \limits_{j=1}^{\infty}(\frac{1}{E_{0\uparrow
}-\hat{H}_{0}}\hat{H_{I}})^{j}\left \vert \downarrow \right \rangle _{R},
\end{equation}
Finally, with the help of Eq.\ref{zero} and Eq.\ref{one}, one can get the
eigenstates of Hamiltonian $\hat{H}$
\[
\left \vert \Phi_{\uparrow}\right \rangle _{R}=\frac{\left \vert \uparrow
\right \rangle _{R}+\sum_{j=1}^{\infty}(\frac{\hat{H_{I}}}{E_{0\uparrow}%
-\hat{H_{0}}})^{j}\left \vert \uparrow \right \rangle _{R}}{{}_{L}\left \langle
\uparrow \right \vert \hat{U_{I}}(0,-\infty)\left \vert \uparrow \right \rangle
_{R}},
\]%
\[
\left \vert \Phi_{\downarrow}\right \rangle _{R}=\frac{\left \vert \downarrow
\right \rangle _{R}+\sum_{j=1}^{\infty}(\frac{\hat{H_{I}}}{E_{0\uparrow}%
-\hat{H_{0}}})^{j}\left \vert \downarrow \right \rangle _{R}}{{}_{L}\left \langle
\downarrow \right \vert \hat{U_{I}}(0,-\infty)\left \vert \downarrow \right \rangle
_{R}}.
\]

To characterize the low energy physics of topologically protected degenerate
ground states, an effective edge Hamiltonian is obtained from the
non-Hermitian higher order perturbation theory,
\begin{equation}
\mathcal{\hat{H}}_{\mathrm{eff}}=\left(
\begin{array}
[c]{cc}%
h_{\uparrow \uparrow} & h_{\uparrow \downarrow}\\
h_{\downarrow \uparrow} & h_{\downarrow \downarrow}%
\end{array}
\right)
\end{equation}
where $h_{\uparrow \uparrow}={}_{L}\left \langle \uparrow \right \vert \hat{H}%
\hat{U}_{I}(0,-\infty)\left \vert \uparrow \right \rangle _{R},$ $h_{\uparrow
\downarrow}={}_{L}\left \langle \uparrow \right \vert \hat{H}\hat{U}%
_{I}(0,-\infty)\left \vert \downarrow \right \rangle _{R},$ $h_{\downarrow
\uparrow}={}_{L}\left \langle \downarrow \right \vert \hat{H}\hat{U}%
_{I}(0,-\infty)\left \vert \uparrow \right \rangle _{R},$ and $h_{\downarrow
\downarrow}={}_{L}\left \langle \downarrow \right \vert \hat{H}\hat{U}%
_{I}(0,-\infty)\left \vert \downarrow \right \rangle _{R} $.

\subsection{Effective non-Hermitian pseudo-spin Hamiltonian}

Firstly, we study the effective pseudo-spin model for topologically degenerate
ground states on an $e\ast o$ lattice that are characterized by topological
closed dynamic strings around the torus $D_{a}(\mathcal{C}_{X-Y}),$
$D_{f}(\mathcal{C}_{Y})$ ($a=v$ or $c$). Here, $\mathcal{C}_{X-Y}$ denotes a
dynamic string along $\hat{e}_{x}-\hat{e}_{y}$-direction and $\mathcal{C}_{Y}$
denotes a dynamic string along $\hat{e}_{y}$-direction, respectively. Due to
$\left \{  D_{a}(\mathcal{C}_{X-Y}),\text{ }D_{f}(\mathcal{C}_{Y})\right \}
=0,$ $D_{a}(\mathcal{C}_{X-Y})$ and $D_{f}(\mathcal{C}_{Y})$ can be mapped
into pseudo-spin operators $\tau^{x}$ and $\tau^{z}$, respectively. Now we map
the two-fold topologically degenerate ground states $\left \vert
m=0\right \rangle $ and $\left \vert m=1\right \rangle $ onto quantum states of
the pseudo-spin $\tau^{z}$ as $\mid \uparrow \rangle$ and $\left \vert
\downarrow \right \rangle $, respectively. Here, $m$ is the number of $\pi$-flux
inside the holes of torus.

Under the perturbation, $\hat{H}_{\mathrm{PT}}$, there are two types of
quantum tunneling processes - virtual e-particle/m-particle propagating along
$\hat{e}_{x}-\hat{e}_{y}$ directions around the torus and virtual fermion
propagating along $\hat{e}_{y}$ direction around the torus.

For the virtual e-particle/m-particle propagating along $\hat{e}_{x}-\hat
{e}_{y}$ directions around the torus, the energy splitting $\Delta \sim{}%
_{L}\langle \uparrow \mid D_{c}(\mathcal{C}_{L_{0}}^{\mathrm{close,topo}%
})\left \vert \downarrow \right \rangle _{R}\sim{}_{L}\langle \downarrow \mid
D_{c}(\mathcal{C}_{L_{0}}^{\mathrm{close,topo}})\left \vert \uparrow
\right \rangle _{R}$ can be obtained by the high-order degenerate-state
perturbation theory as
\begin{equation}
\Delta=2L_{x}L_{y}\frac{(h_{x})^{L_{0}}}{(-4g)^{L_{0}-1}},
\end{equation}
where the step number $L_{0}$ is equal to
\begin{equation}
\frac{L_{x}L_{y}}{\xi}.
\end{equation}
Here $\xi$ is the maximum common divisor for $L_{x}$ and $L_{y}$. Because the
topological closed dynamic string $D_{c}(\mathcal{C}_{L_{0}}%
^{\mathrm{close,topo}})$\ (or $D_{v}(\mathcal{C}_{L_{0}}^{\mathrm{close,topo}%
})$) of e-particle (or m-particle) plays a role of $\tau^{x}$ on the quantum
states, we obtain the effective pseudo-spin Hamiltonian for topologically
degenerate ground states due to the contribution of e-particle (or m-particle)
as
\begin{equation}
\frac{\Delta}{2}\tau^{x};
\end{equation}
For the tunneling process of f-particle propagating around the torus along
direction $\hat{e}_{y}$, we obtain the energy difference
\begin{equation}
\varepsilon \sim{}_{L}\langle \uparrow \mid D_{f}(\mathcal{C}_{L_{y}%
}^{\mathrm{close,topo}})\left \vert \uparrow \right \rangle _{R}\sim-{}%
_{L}\langle \downarrow \mid D_{f}(\mathcal{C}_{L_{y}}^{\mathrm{close,topo}%
})\left \vert \downarrow \right \rangle _{R}%
\end{equation}
of topologically degenerate ground states as
\begin{equation}
\varepsilon=16L_{x}L_{y}g(\frac{ih_{z}}{8g})^{L_{y}}.
\end{equation}
Because the topological closed dynamic string $D_{f}(\mathcal{C}_{L_{y}%
}^{\mathrm{close,topo}})$\ of f-particles plays a role of $\tau^{z}$ on the
quantum states, we obtain the effective pseudo-spin Hamiltonian for
topologically degenerate ground states due to the contribution of f-particles
as
\begin{equation}
\frac{\varepsilon}{2}\tau^{z}.
\end{equation}

Finally, the effective pseudo-spin Hamiltonian of the two topologically
degenerate ground states on an $e\ast o$ lattice is obtained as
\begin{equation}
\mathcal{\hat{H}}_{\mathrm{eff}}^{e\ast o}=\frac{\Delta}{2}\tau^{x}%
+\frac{\varepsilon}{2}\tau^{z}=\frac{1}{2}\left(
\begin{array}
[c]{cc}%
\beta(ih_{z})^{L_{y}} & \alpha h_{x}^{L_{0}}\\
\alpha h_{x}^{L_{0}} & -\beta(ih_{z})^{L_{y}}%
\end{array}
\right)
\end{equation}
where
\begin{equation}
\Delta=\alpha(h_{x})^{L_{0}},
\end{equation}
and
\begin{equation}
\varepsilon=\beta(ih_{z})^{L_{y}}.
\end{equation}
$\alpha$ and $\beta$ are two real parameters.

According to odd number $L_{y},$
\begin{equation}
\operatorname{Re}\varepsilon \equiv0,
\end{equation}
and
\begin{equation}
\operatorname{Im}\varepsilon=16L_{x}L_{y}g(\frac{h_{z}}{8g})^{L_{y}}\neq0,
\end{equation}
the effective pseudo-spin model $\mathcal{\hat{H}}_{\mathrm{eff}}$ turns into
the typical $\mathcal{PT}$ symmetric non-Hermitian Hamiltonian that is
invariant under a combined parity ($\mathcal{P}$) and time-reversal
($\mathcal{T}$) symmetry for $h_{x}$ and $g$, where $\mathcal{P}%
\rightarrow \tau^{x}$ simply corresponds to an exchange of the two
topologically degenerate ground states and $\mathcal{T}:i\rightarrow-i$. As a
result, for odd number $L_{y}$, we have
\begin{equation}
\left[  \mathcal{P},\mathcal{\hat{H}}_{\mathrm{eff}}^{e\ast o}\right]
\neq0\text{ and }\left[  \mathcal{T},\mathcal{\hat{H}}_{\mathrm{eff}}^{e\ast
o}\right]  \neq0,
\end{equation}
but
\begin{equation}
\left[  \mathcal{PT},\text{ }\mathcal{\hat{H}}_{\mathrm{eff}}^{e\ast
o}\right]  =0.
\end{equation}

We may use similar approach to get the effective non-Hermitian pseudo-spin
model for topologically degenerate ground states on an $o\ast e$ lattice
($\mathcal{\hat{H}}_{\mathrm{eff}}^{o\ast e}$) and that on an $o\ast o$
lattice ($\mathcal{\hat{H}}_{\mathrm{eff}}^{o\ast o}$) with different $\Delta$
and $\varepsilon$.

\subsection{$\mathcal{PT}$-symmetry spontaneous breaking and anomalous
topological degeneracy}

Next, we focus on the topologically degenerate ground states for $\hat
{H}_{\mathrm{NTO}}$ on an $e\ast o$ lattice and those on $o\ast o$ lattice. An
interesting phenomenon is $\mathcal{PT}$-symmetry spontaneous breaking.

The effective pseudo-spin model for topologically degenerate ground states on
an $e\ast o$ lattice and those on $o\ast o$ lattice are obtained as
\begin{equation}
\mathcal{\hat{H}}_{\mathrm{eff}}=\frac{\Delta}{2}\tau^{x}+\frac{\varepsilon
}{2}\tau^{z}.
\end{equation}
The eigenvalues and (non-normalized) eigenvectors of $\mathcal{\hat{H}%
}_{\mathrm{eff}}$ are
\begin{equation}
E_{\pm}=\pm \frac{1}{2}\sqrt{\Delta^{2}-\left \vert \varepsilon \right \vert ^{2}%
},
\end{equation}
and
\begin{align}
\left \vert \psi_{+}\right \rangle  &  =e^{i\frac{\Theta}{2}}\left \vert
\uparrow \right \rangle _{R}+e^{-i\frac{\Theta}{2}}\left \vert \downarrow
\right \rangle _{R},\nonumber \\
\left \vert \psi_{-}\right \rangle  &  =ie^{-i\frac{\Theta}{2}}\left \vert
\uparrow \right \rangle _{R}-ie^{i\frac{\Theta}{2}}\left \vert \downarrow
\right \rangle _{R},
\end{align}
where $\Delta$ is real parameter, $\varepsilon$ is imaginary parameter, and
$\sin \Theta=\left \vert \varepsilon \right \vert /\Delta$\cite{Ramezani}. The
energy splitting of two topologically degenerate ground states is $\delta
E=E_{+}-E_{-}$. For the case of $|\Delta| \geq \left \vert \varepsilon
\right \vert $ the system belongs to a phase with $\mathcal{PT}$ symmetry, of
which $E_{+}$ and $E_{-}$ are real and the eigenvectors are eigenstates of the
symmetry operator, i.e.,
\begin{equation}
\mathcal{PT}\left \vert \psi_{\pm}\right \rangle =\left \vert \psi_{\pm
}\right \rangle .
\end{equation}
For the case of $|\Delta|<\left \vert \varepsilon \right \vert $, $E_{+}$ and
$E_{-}$ are imaginary that correspond to a gain eigenstate and a loss
eigenstate, respectively, i.e.,
\begin{equation}
\mathcal{PT}\left \vert \psi_{\pm}\right \rangle \neq \left \vert \psi_{\pm
}\right \rangle .
\end{equation}
A $\mathcal{PT}$-symmetry-breaking transition occurs at the exceptional points
$\left \vert \varepsilon \right \vert =|\Delta|$ that leads to the following
relation
\begin{equation}
\alpha(h_{x})^{\frac{L_{x}L_{y}}{\xi}}=\beta(h_{z})^{L_{y}}.
\end{equation}
\emph{Anomalous topological degeneracy} occurs, i.e., the number of the
topologically protected ground states is reduced from $2$ to $1$.

\begin{figure}[ptb]
\includegraphics[clip,width=0.35\textwidth]{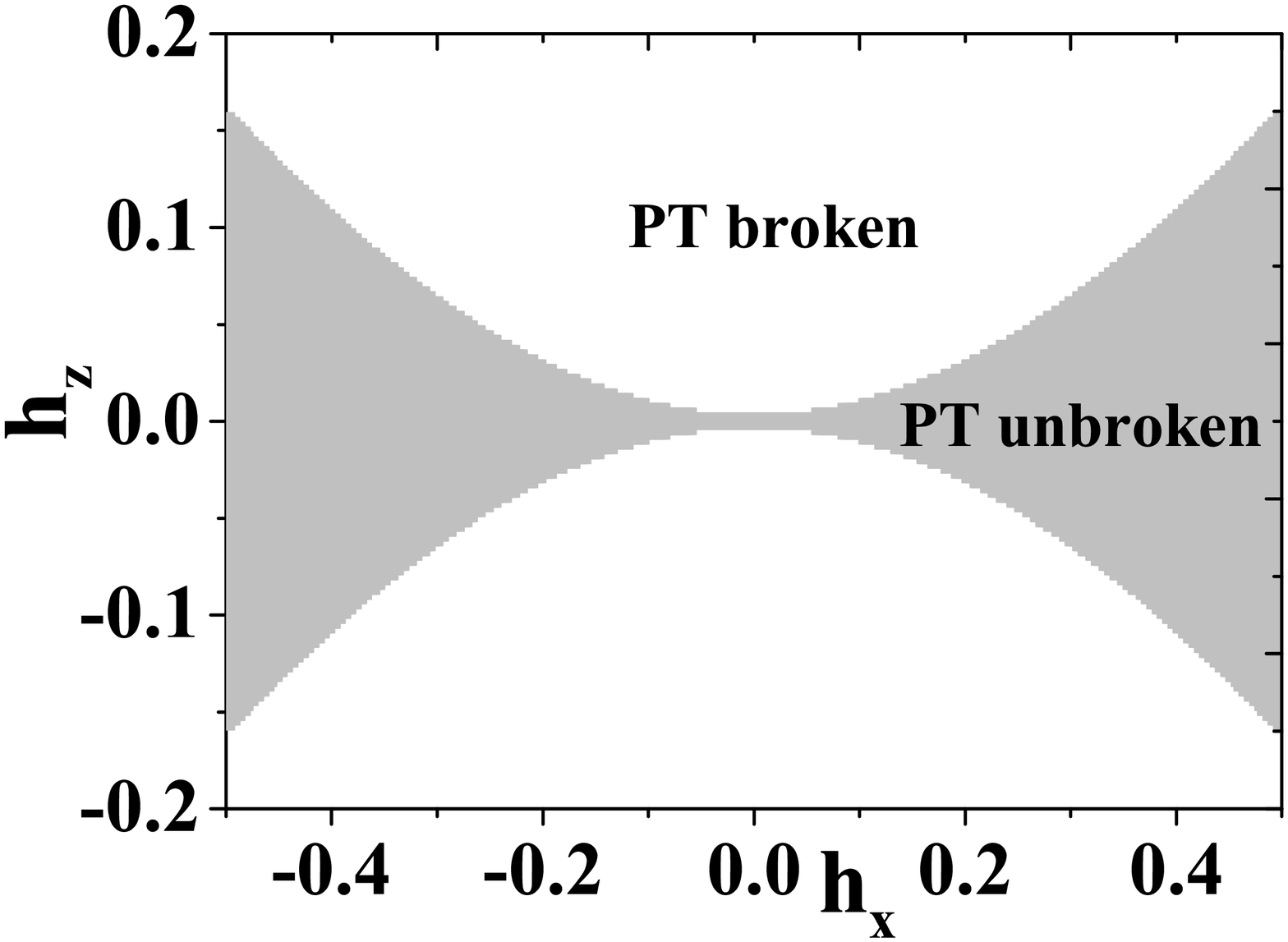}\caption{(Color online)
The phase diagram of $\mathcal{PT}$-symmetry spontaneous breaking for the
topologically degenerate ground states on $2\ast3$ lattice. The phase boundary
is exceptional points.}%
\end{figure}\begin{figure}[ptb]
\includegraphics[clip,width=0.45\textwidth]{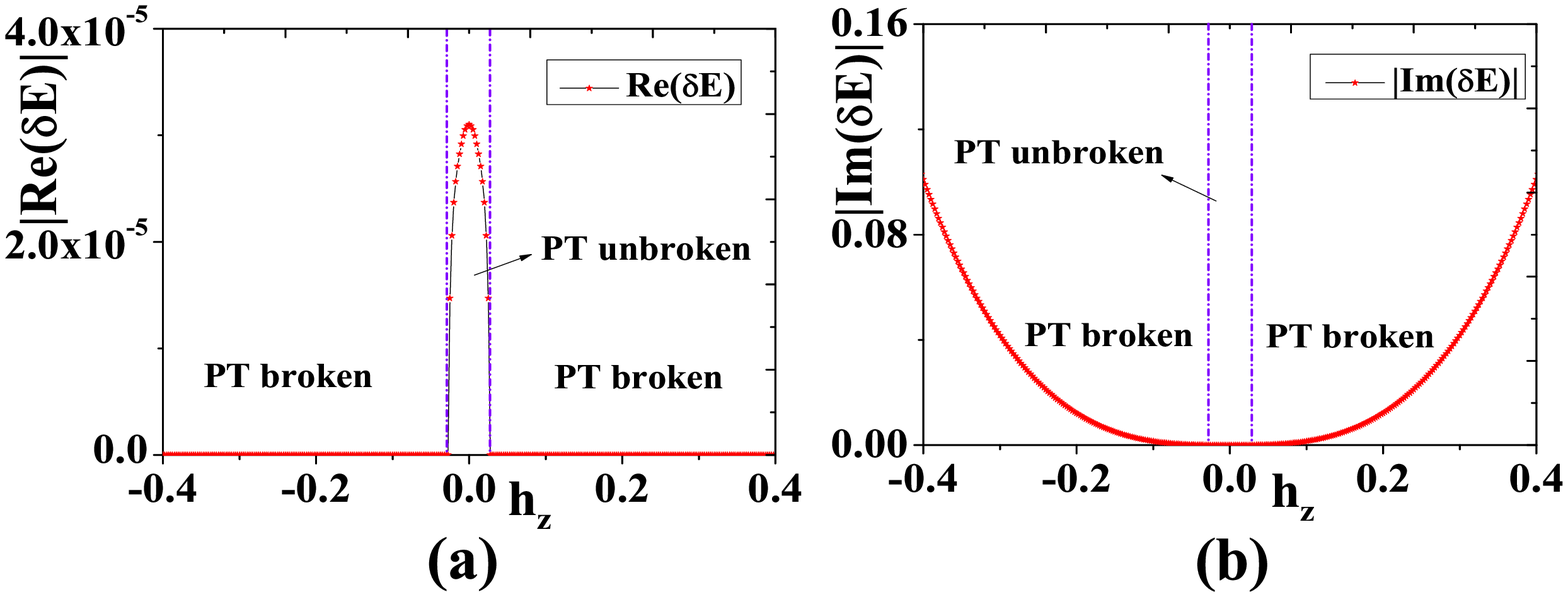}\caption{(Color online)
(a)(b) The absolute value of real part and imaginary of energy splitting for
the two degenerate ground states for the case of $h_{x}=0.2$ via $h_{z}$ based
on the non-Hermitian Wen-plaquette model on $2\ast3$ lattice.}%
\end{figure}\begin{figure}[ptb]
\includegraphics[clip,width=0.35\textwidth]{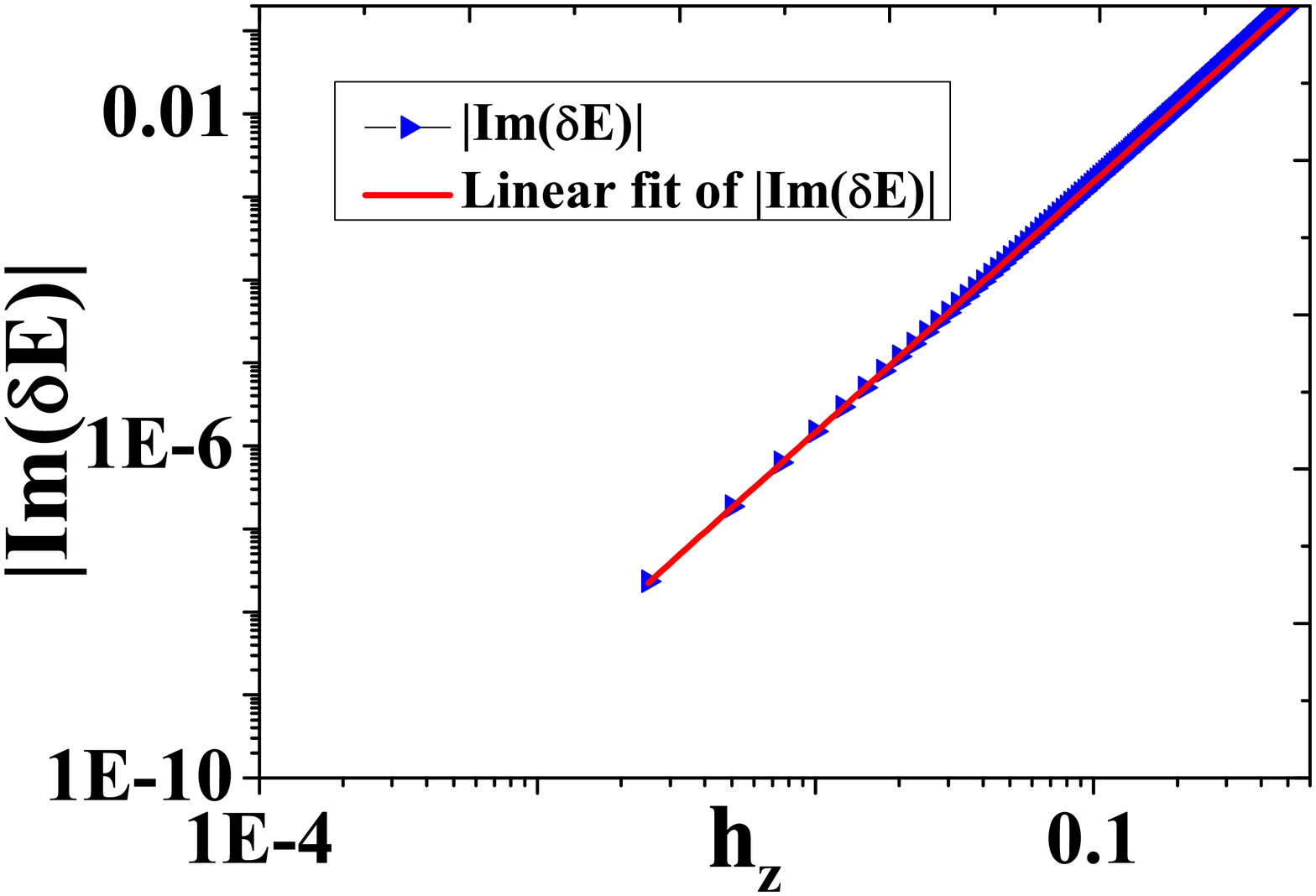}\caption{(Color online)
The absolute value of imaginary parts of energy splitting between the two
degenerate ground states $\left \vert \operatorname{Im}(\delta E)\right \vert $
for the case of $h_{x}=0$ via $h_{z}$ on $2\ast3$ lattice.}%
\end{figure}

In Fig.7, Fig.8 and Fig.9, we plot the numerical results from the exact
diagonalization technique of the non-Hermitian Wen-plaquette model $\hat
{H}_{\mathrm{NTO}}$ on $2\ast3$ lattice with periodic boundary conditions.
Fig.7 shows the global phase diagram of $\mathcal{PT}$-symmetry-breaking
transition for topologically degenerate ground states. The phase boundary are
all exceptional points characterized by the relation $\alpha(h_{x})^{6}%
=\beta(h_{z})^{3}$. In Fig.8(a) and Fig.8(b), we plot the absolute value of
real part ($\left \vert \operatorname{Re}(\delta E)\right \vert $) and imaginary
($\left \vert \operatorname{Im}(\delta E)\right \vert $) of energy splitting for
the two degenerate ground states for the non-Hermitian Wen-plaquette model
$\hat{H}_{\mathrm{NTO}}$ with $h_{x}=0.2$ on $2\ast3$ lattice, respectively.
The results ($\left \vert \operatorname{Im}(\delta E)\right \vert $) in Fig.9
are consistent to theoretical prediction: the step of dynamic strings for
fermions is $3.01$ from the numerical results (the theoretical prediction is
$3$).

\begin{figure}[ptb]
\includegraphics[clip,width=0.35\textwidth]{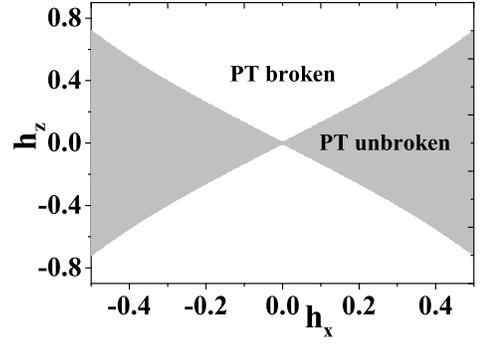}\caption{(Color online)
The phase diagram of $\mathcal{PT}$-symmetry spontaneous breaking for the
topologically degenerate ground states on $3\ast3$ lattice. The phase boundary
is exceptional points.}%
\end{figure}\begin{figure}[ptb]
\includegraphics[clip,width=0.5\textwidth]{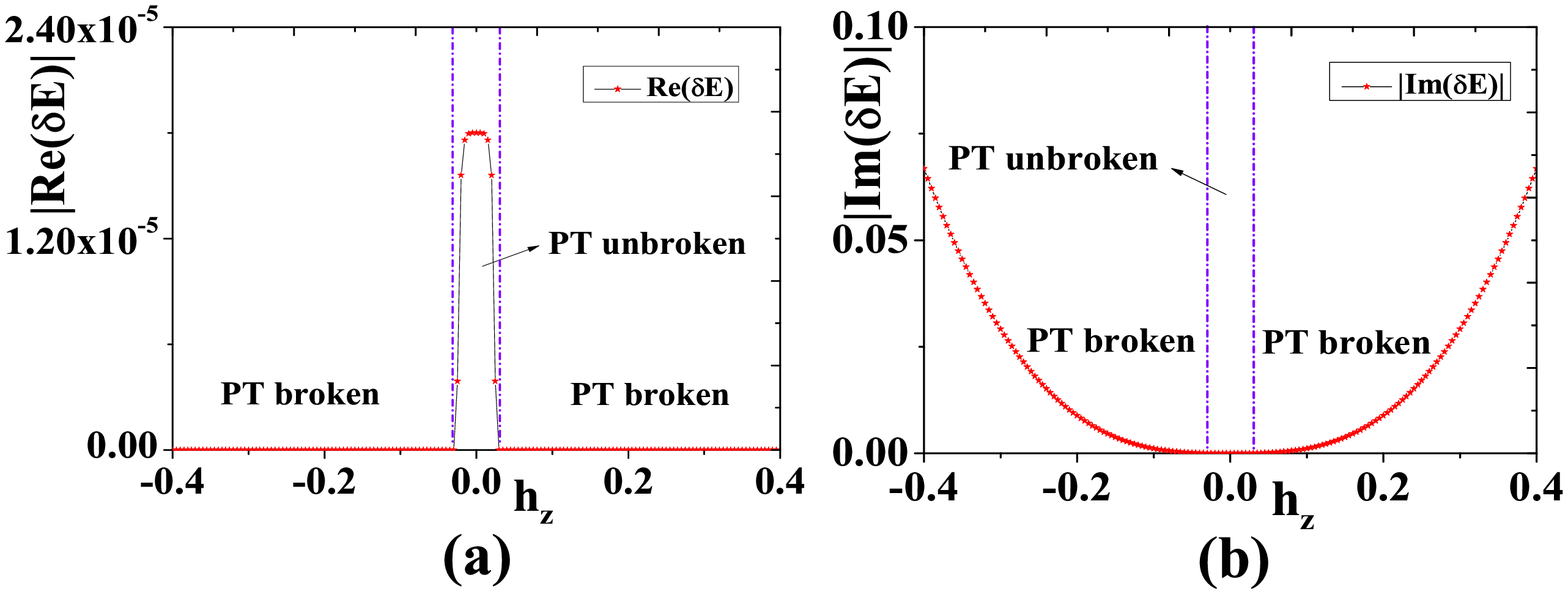}\caption{(Color online)
(a)(b) The absolute value of real part and imaginary of energy splitting for
the two degenerate ground states for the case of $h_{x}=0.02$ via $h_{z}$
based on the non-Hermitian Wen-plaquette model on $3\ast3$ lattice.}%
\end{figure}\begin{figure}[ptb]
\includegraphics[clip,width=0.35\textwidth]{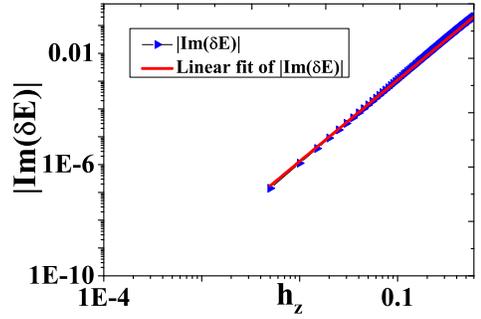}\caption{(Color online)
The absolute value of imaginary parts of energy splitting between the two
degenerate ground states $\left \vert \operatorname{Im}(\delta E)\right \vert $
for the case of $h_{x}=0$ via $h_{z}$ on $3\ast3$ lattice.}%
\end{figure}

In Fig.10, Fig.11 and Fig.12, we plot the numerical results from the exact
diagonalization technique of the non-Hermitian Wen-plaquette model $\hat
{H}_{\mathrm{NTO}}$ on $3\ast3$ lattice with periodic boundary conditions.
Fig.10 shows the global phase diagram of $\mathcal{PT}$-symmetry-breaking
transition for topologically degenerate ground states. The phase boundary are
exceptional points characterized by the relation $h_{z}\sim h_{x}$, which is
agreement with the result from pseudo-spin model above $\alpha h_{x}^{3}%
=\beta(h_{z})^{3}$. In Fig.11(a) and Fig.11(b), we plot the absolute value of
real parts and imaginary parts of energy splitting for the two topologically
degenerate ground states for the non-Hermitian Wen-plaquette model with
$h_{x}=0.02$ on $3\ast3$ lattice, respectively. In Fig.12, the absolute value
of imaginary parts of energy splitting between the two degenerate ground
states of the non-Hermitian Wen-plaquette (toric-code) model is obtained with
$h_{x}=0$ on $3\ast3$ lattice. The numerical result indicates that the step of
dynamic strings for fermions is $2.92$, which is consistent with theoretical
prediction $3$.

\subsection{Application}

It was pointed out that the topological degenerate ground states for a $Z_{2}$
topological order make up a protected subspace free from
error\cite{k1,k2,kou1,kou2,kou3}. One can manipulate the protected subspace by
controlling their quantum tunneling effect\cite{kou1,kou2,kou3,md1}. However,
due to the very tiny value of energy splitting for topological degenerate
ground states, the initialization topological degenerate ground states for
quantum computation based on such topological qubits becomes very difficult.
As a result, we propose a new approach to initialize the system (for example,
the topological degenerate ground states on an $e\ast o$ lattice) into a
particular state. The basic idea is \emph{to drive the topological qubit to
exceptional points by adding (real or imaginary) external field and then
removal it slowly}.

The evolution will occur along the exceptional points according to the
Hamiltonian
\begin{equation}
\hat{H}_{\mathrm{NTO}}(t)=\hat{H}_{\mathrm{TO}}+%
%TCIMACRO{\dsum \limits_{i}}%
%BeginExpansion
{\displaystyle \sum \limits_{i}}
%EndExpansion
[h_{x}(t)\sigma_{i}^{x}+ih_{z}(t)\sigma_{i}^{z}],
\end{equation}
where
\begin{equation}
h_{x}{(t)={h}_{0}(1-e}^{t/t_{0}}),
\end{equation}
and
\begin{equation}
h_{z}(t)=(\frac{\alpha}{\beta})^{\frac{1}{L_{y}}}(h_{x}(t))^{\frac{L_{x}}{\xi
}}.
\end{equation}
At the beginning, $t\rightarrow-\infty,$ under the finite external field
${h}_{x}{(t),}$ ${h_{z}(t),}$ the system is at exceptional points. The
effective Hamiltonian of the topological qubit in the external field
\begin{equation}
\mathcal{\hat{H}}_{\mathrm{eff}}^{e\ast o}(t)=\frac{\Delta(t)}{2}(\tau
^{x}+i\tau^{z}).
\end{equation}
Due to the non-Hermitian term, the topological degeneracy is reduced into
non-Hermitian degeneracy at the exceptional points. As a result, there doesn't
exist topological degeneracy and the two topological degenerate ground states
merge into one quantum "steady" state,
\begin{equation}
\left \vert \psi_{+}\right \rangle \rightarrow \left \vert \uparrow \right \rangle
_{R}-i\left \vert \downarrow \right \rangle _{R}\Leftrightarrow \left \vert
\psi_{-}\right \rangle \rightarrow \left \vert \uparrow \right \rangle
_{R}-i\left \vert \downarrow \right \rangle _{R}.
\end{equation}
At the time $t=0,$ when the external field disappears, ${h_{x}(t),}$
${h_{z}(t)=0}$, a pure quantum state for quantum computation based on
topological qubits is initialized as
\begin{equation}
\left \vert \uparrow \right \rangle _{R}-i\left \vert \downarrow \right \rangle
_{R}.
\end{equation}
This approach to initialization for topological qubits is much more efficiency
than the traditional approach by adding a (pure) real external field in
Ref.\cite{kou1,kou2,kou3}.

\section{Conclusion}

In this paper, we developed a theory of the non-Hermitian $Z_{2}$ topological
order. The quantum states of non-Hermitian $Z_{2}$ topological order are
characterized by different Hermitian/non-Hermitian dynamic strings. The ground
states of non-Hermitian Wen-plaquette model on a torus are classified by
different Hermitian/non-Hermitian topological closed dynamic strings. The
effective model for the topologically degenerate ground states on $e\ast o$,
or $o\ast e,$ or $o\ast o$ lattices can be mapped onto an effective
non-Hermitian pseudo-spin model, from which the anomalous topological
degeneracy occurs. In particular, there exists spontaneous $\mathcal{PT}$
symmetry breaking for the topologically degenerate ground states. At
\textquotedblleft exceptional points\textquotedblright, the topologically
degenerate ground states merge and the topological degeneracy turns into
non-Hermitian degeneracy. In the end, the application of the non-Hermitian
$Z_{2}$ topological order and its possible physics realization are discussed.
In addition, this work may help people to understanding the effect of
non-Hermitian perturbations on many-body systems.

In the end, we address the experimental realization of the non-Hermitian
$Z_{2}$ topological order. It is still of challenge \emph{both} to
experimentally investigate $\mathcal{PT}$ symmetric Hamiltonian related
physics in quantum systems and to realize the toric-code model. A possible
approach is cold-atom experiments: On the one hand, non-Hermitian Hamiltonians
arise in cold-atom experiments due to spontaneous decay\cite{Hang2013,
Lee2014,luo}; On the other hand, a small system for toric-code model with open
boundary condition has also been realized in cold atoms\cite{yuan}.

\acknowledgments This work is supported by NSFC Grant No. 11674026, 11974053.
We thank Xiao-Gang Wen for helpful discussion.

\end{document}